\documentclass[apJ]{emulateapj}
\usepackage{textcomp}
\usepackage{epsfig, epstopdf}
\usepackage{amssymb,amsmath}
\usepackage{longtable,lscape}
\usepackage[]{natbib}
\newcommand{\lnh}{\hbox{log~$N_{\rm H}$}}
\newcommand{\lnhcm}{\hbox{log~($N_{\rm H}/{\rm cm}^{-2})$}}

\newcommand{\lLa}{\hbox{log~$L_{2-10}$}}
\newcommand{\lip}{\hbox{log~$\xi$}}
\newcommand{\lU}{\hbox{log~$U$}}
\newcommand{\kms}{\hbox{km~s$^{-1}$}}
\newcommand{\cmsq}{\hbox{cm$^{-2}$}}
\newcommand{\cc}{\hbox{cm$^{-3}$}}

\newcommand{\lumin}{\hbox{erg~s$^{-1}$}}

\newcommand{\aox}{\hbox{$\alpha_{\rm ox}$}}
\newcommand{\nh}{\hbox{${N}_{\rm H}$}}
\newcommand{\nhat}{\hbox{${\hat{N}}_{\rm H}$}}
\newcommand{\den}{\hbox{${n}_{\rm H}$}}
\newcommand{\lden}{\rm log~\den}
\newcommand{\apm}{\hbox{APM~08279+5255}}
\newcommand{\pg}{\hbox{PG~1115+080}}

\newcommand{\OF}{\hbox{observed-frame}}
\newcommand{\RF}{\hbox{rest-frame}}

\newcommand{\FeXXV}{\hbox{Fe {\sc xxv}}}
\newcommand{\FeXXVI}{\hbox{Fe {\sc xxvi}}}

\newcommand{\CIV}{\hbox{C {\sc iv}}}

\newcommand{\NV}{N {\sc v}}
\newcommand{\OVI}{O {\sc vi}}

\newcommand{\XR}{\hbox{X-ray}}
\newcommand{\chandra}{\emph{Chandra}}
\newcommand{\suzaku}{\emph{Suzaku}}

\newcommand{\xmm}{\hbox{\emph{XMM-Newton}}}

\newcommand{\Am}{\hbox{\tiny \AA}}

\newcommand{\PL}{\hbox{power-law}}
\newcommand{\xspec}{{\sc xspec}}
\newcommand{\xstar}{{\sc xstar}}
\newcommand{\cloudy}{{\sc cloudy}}
\newcommand{\xscort}{{\sc xscort}}
\newcommand{\MF}{\hbox{Mathews-Ferland}}

\begin{document}

\title{A STUDY OF THE X-RAYED OUTFLOW OF \apm\ THROUGH PHOTOIONIZATION CODES.}

\author{Cristian Saez,\altaffilmark{1} \& George Chartas\altaffilmark{2}}

\altaffiltext{1}{Department of Astronomy \& Astrophysics, Pennsylvania State University,
University Park, PA 16802, saez@astro.psu.edu}

\altaffiltext{2}{Department of Physics and Astronomy, College of Charleston, Charleston, SC, 29424, chartasg@cofc.edu}

\keywords{cosmology: observations
--- X-rays: galaxies --- galaxies: active --- quasars: absorption lines}

\begin{abstract}

\setcounter{footnote}{2}

We present new results from our study of the X-rayed outflow of  the $z = 3.91$ gravitationally lensed broad absorption line (BAL) quasar APM 08279+5255. These  results are based on spectral fits to all the long exposure observations of \apm\ using a new quasar-outflow model. This model is based on \cloudy \footnote{\cloudy\ is a photoionization code designed to simulate conditions in interstellar matter under a broad range of conditions. We have used version 08.00 of the code last described by \cite{Fer98}. The atomic database used by \cloudy\ is described in  \cite{Fer01} and http://www.pa.uky.edu/$\sim$verner/atom.html.} simulations of  a near-relativistic quasar outflow. 

The main conclusions from our multi-epoch spectral re-analysis of \chandra,  \xmm\ and \suzaku\ observations of \apm\ are: 1) In every observation we confirm the presence of two strong features, one at rest-frame energies between 1--4~keV, and the other between 7--18~keV.  2) We confirm that the low-energy absorption (1--4~keV rest-frame)  arises from a low-ionization absorber with $\lnhcm \sim 23$ and the high-energy absorption (7$-$18 keV rest-frame) arises from highly ionized ($3 \lesssim \lip \lesssim  4$; where $\xi$ is the ionization parameter) iron in a near-relativistic outflowing wind. Assuming this interpretation, we find that the velocities on the outflow could get up to $\sim$~0.7c.   3) We confirm a correlation between the maximum outflow velocity and the photon index and find possible trends between the maximum outflow velocity and the X-ray luminosity, and between the total column density and the photon index.

We performed calculations of the force multipliers of material illuminated by absorbed power laws and a Mathews-Ferland SED. 
We found that variations of the X-ray and UV parts of the SEDs and the presence of a moderate absorbing shield will produce important changes in the strength of the radiative driving force. These results support the observed trend found between the outflow velocity and X-ray photon index in APM 08279+5255. If this result is confirmed it will imply that radiation pressure is an important mechanism in producing quasar outflows.
\end{abstract}

\section{INTRODUCTION}

The existence of a $M_{\rm BH}-\sigma$ relation in nearby galaxies \citep[e.g.,][]{Fer00} suggests that a feedback mechanism exists regulating the co-evolution between the massive black hole at the center of a galaxy and the 
formation of its bulge. Quasar outflows could transport a fraction of the central 
black hole binding energy to the host galaxy by possibly shocking against the host 
interstellar medium (ISM) and therefore may be responsible for the $M_{\rm BH}-\sigma$ relation \citep[e.g.,][]{Kin11}.
Fast and powerful AGN jets and winds can be formed mainly by magnetic or radiative driving. 
Magnetically driven jets are thought to be responsible for the production of bubbles and cavities in galaxy clusters. 
The sources of these jets are found to be radio-loud (RL) AGNs located in the central giant elliptical galaxies \citep[see][and references therein]{McN07}. 
Although there is compelling evidence for the existence of jets in RL AGNs \citep[see e.g.,][and references therein]{Bri84}, 
the presence of jets or magnetically-driven winds in radio quiet AGNs is still debated.
The mechanism driving winds in the majority of AGNs therefore is currently uncertain since the majority of AGN are radio quiet.\footnote{RL AGNs represent between 10-20\% of the whole AGN population \citep[e.g.,][]{Ive02, Jia07}} It is important to note, however,  that a recent study claims observational evidence of magnetically-driven winds in high signal-to-noise ratio (S/N)  spectra of nearby AGNs \citep[][]{Fuk10}. 
Specifically, \cite{Fuk10} find that a MHD model can explain the distribution of absorption as a function of ionization parameter ($d \lnh/d\xi$) in the X-ray spectra of 
five nearby AGNs (mostly Seyfert 1 galaxies at $z \lesssim 0.1$). In is not clear if this result can be extended to high redshift sources or sources with   
relativistic winds where we expect strong and likely discontinuous gradients of the outflow velocity and therefore we do not expect
to find smooth ${\rm d} \lnh / d \xi$ distributions as required in their model.

At moderate to high Eddington ratios ($L/ L_{\rm Edd} \gtrsim 0.2$), which may be common in high redshift sources $z \gtrsim 2$ \citep{Kol06}, we expect that radiative driving becomes more important than magnetic driving \citep{Eve05}.  Recent observations of the high redshift ($z \approx 2$) ultraluminous infrared galaxy (ULIRG) SMM J1237+6203 suggest a large-scale powerful wind in this object driven by AGN radiation \citep{Ale10}. These observations are suggestive that in high redshift AGNs  ($z \gtrsim 2$) radiatively driven winds may be more common than the jets found in radio galaxies.  Radiative driving is also the favorite mechanism for explaining the formation of outflows observed in the UV from BAL quasars. Evidence of the later is found in the correlation between changes in the maximum velocity of the outflow (\CIV\  absorption lines) as a function of variations in the soft-to-hard spectral slope \aox\   \citep[e.g.,][]{Gib09}. The outflows observed in the UV from BAL quasars usually have moderate velocities ($ \lesssim 0.1 c$),  however,  there have been reported cases with velocities of $\sim 0.2 c$ \citep{Rod07}. Observations \citep{Ham08} of the quasar J105400.40+034801.2 imply wind velocities of $\sim 0.8 c$ based on measurements of the \CIV\ line. We note, however, that even though this source reveals blue-shifted broadenings $\gtrsim 4000~\kms$ in \CIV\ lines, it is not considered a BAL quasar in the strict sense of the definition \citep[i.e.][]{Wey91}.\footnote{The commonly used definition of BAL quasar \citep[i.e.][]{Wey91} does not consider outflow velocities $> 25,000~\kms$.
BAL winds have also been found in the X-ray band \citep{Cha02, Cha07b}, however, the X-ray faintness of BAL quasars \citep{Gal06} has limited the number of cases with observed X-ray BALs to a few.} 
Several of the reported  low S/N cases of high velocity AGN outflows  have led to spurious detections \citep[see][and references therein]{Vau08}.
A handful of moderate S/N X-ray spectra of BAL and mini-BAL quasars have been obtained that show strong signatures of fast winds
 \citep[e.g.,][]{Cha02, Cha07a, Cha07b, Cha09a}. The velocities of the outflowing X-ray absorbing material in some cases are found to be ($\gtrsim 0.7c$) indicating a different dynamical state than the observed outflow in the UV band. The fast variability and high ionization state of the winds observed in the X-rays  could also be indicating that these X-ray outflows 
 are originating closer to the central source than the outflows observed in the UV.

Recently \citep{Sae09, Cha09a} we presented observational evidence for the presence of a powerful outflow 
in BAL quasar \apm\  based on the analysis of the X-ray spectra of this object.
APM~08279+5255 is  a high redshift quasar which is unusually bright due to gravitational lensing, providing the exceptional opportunity to study it with high-quality spectral data. Our conclusions from these studies were that the outflow was accelerated to relativistic speeds, and was releasing kinetic energy at a rate comparable to the bolometric luminosity of this source. The latter indicated that this outflow can release an important fraction of the energy produced by black-hole accretion into the surrounding galaxy. In our study of \apm\ we also found evidence that the outflow could be radiatively accelerated by the central source.  \newline This paper is laid out as follows. In \S\ref{S:outo} we present the main conclusions extracted from recent studies that focused on long X-ray observations ($\sim$100 ks) of \apm\  performed with \chandra,  \xmm, and \suzaku.  We also expand on the previous studies in two directions. First in \S\ref{S:nro} we describe how the use of photoionization models may help us  better constrain the  physical properties of the outflowing \XR\ absorbing gas.  Second in \S\ref{S:SEDs} we discuss how the SED of the central source may influence the  dynamics of the outflow. 

Unless stated
otherwise, throughout this paper we use CGS units, the errors
listed are at the 1-$\sigma$ level, and we adopt a flat
$\Lambda$-dominated universe with $H_0=70~\kms$Mpc$^{-1}$,
$\Omega_\Lambda=0.7$, and $\Omega_M=0.3$.

\section{The fast outflow of  \apm} \label{S:outo}

This section contains information extracted from past studies of \apm. In \S \ref{S:gapm}, we concentrate on general information that is relevant to understanding the origin of the fast outflow of X-ray absorbing material observed in this object. 
In (\S \ref{S:Xobs}), we describe the most important results extracted from eight deep 
X-ray observations of  APM 08279+5255.

\begin{figure}
\includegraphics[width=8.6cm]{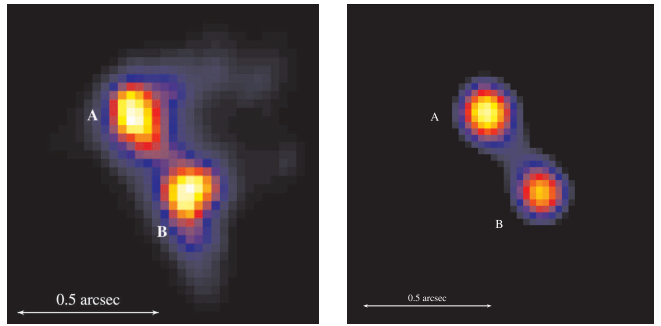}
  \centering
\caption[Deconvolved image of the 2002 February 24 Chandra observation of \apm.]{Deconvolved image (left panel) of the 2002 February 24 \chandra\  observation of \apm. North is up, and east is to the left. Best fit PSF model (right panel) to the same observation of \apm. }
\label{fig:iapm}
\end{figure}

\subsection{Properties of the fast outflow of \apm} \label{S:gapm}

APM~08279+5255 is a gravitationally lensed BAL quasar at redshift z=3.911 \citep{Dow99}.  Its bolometric luminosity estimated as $L_{\rm bol} = 7 \times 10^{15} \mu_L^{-1} L_\odot$ \citep[$\mu_L$ is the lens amplification factor;][]{Irw98, Rie09} makes this source one of the most luminous in the universe\footnote{As a reference the most luminous quasars in the SDSS have $L_{\rm bol} \sim 10^{14} L_{\odot}$ \citep[e.g.,][]{Gan07}.}, despite the value of the lens amplification factor.
The presence of high-ionization BAL features produced in \CIV, \OVI, \NV\ and Ly~$\alpha$ transitions indicates that it is a high ionization BAL quasar (HiBAL).  The \CIV\ BAL of \apm\ contains multiple absorption troughs with widths ranging between 2000~\kms\ and 2500~\kms. The maximum observed outflow velocity of the \CIV\ absorption is $v_{\rm max(CIV)} \sim 0.04c$ relative to the source systemic redshift of $z = 3.911$  \citep{Sri00}.


\begin{deluxetable*}{cccccccccccc} \tabletypesize{\scriptsize}
\tablecolumns{12} \tablewidth{0pt} \tablecaption{Log of observations of \apm. \label{tab:obid}}
\tablehead{
 \multicolumn{1}{c}{Date$^a$} & \multicolumn{1}{c}{OBS ID}  & \multicolumn{1}{c}{Telescope} & \multicolumn{1}{c}{Instrument} & \multicolumn{1}{c}{Exposure} & \multicolumn{1}{c}{Net exp} &\multicolumn{1}{c}{Net counts$^b$} &\multicolumn{1}{c}{$f_{0.5-8}$$^c$}   \\
}
\startdata
2002-02-24 (Epoch1) & 2979  & \chandra & ACIS BI & 91.9 ks & 88.8 ks  & 5,627$\pm$75 &   5.44$\pm$0.15 \\
2002-04-28 (Epoch2)& 0092800201  & \xmm & EPIC pn & 102.9 ks & 83.5 ks & 12,820$\pm$139 & 5.51$\pm$0.15 \\ 
2006-10-12  (OBS1) & 701057010 & \suzaku & XIS FI & 102.3 ks & 71.3 ks  & 6,071$\pm$90 &  5.58$\pm$0.34 \\
2006-10-12  (OBS1) & 701057010 & \suzaku & XIS BI & 102.3 ks & 71.3 ks & 2,325$\pm$64 & 5.39$\pm$0.43\\
2006-11-01  (OBS2) & 701057020  & \suzaku & XIS FI & 102.3 ks & 67.9 ks & 5498$\pm$87 &  5.18$\pm$0.32  \\
2006-11-01  (OBS2) & 701057020 & \suzaku & XIS BI & 102.3 ks & 67.9 ks &  2,181$\pm$62 & 5.15$\pm$0.37  \\ 
2007-03-24  (OBS3) & 701057030 & \suzaku & XIS FI & 117.1 ks & 86.4 ks & 4750$\pm$80 & 5.55$\pm$0.49 \\ 
2007-03-24  (OBS3) & 701057030 & \suzaku & XIS BI & 117.1 ks & 86.4 ks & 2,918$\pm$71 &  5.63$\pm$0.39 \\ 
2007-10-06 (Epoch3) & 0502220201& \xmm &   EPIC pn & 89.6 ks & 56.4 ks& 11,400$\pm$114 & 6.04$\pm$0.16 \\ 
2007-10-22 (Epoch4) & 0502220301 & \xmm &    EPIC pn & 90.5 ks & 60.4 ks & 16,698$\pm$133 & 8.06$\pm$0.15 \\ 
2008-01-14 (Epoch5) & 7684 & \chandra & ACIS BI & 91.7 ks & 88.06 ks & 6,938$\pm$83 & 5.89$\pm$0.21 \\
\enddata

\tablenotetext{a}{The date is also described by the term in parentheses.}

\tablenotetext{b}{Background-subtracted source counts including events with energies 
within the 0.2--10 keV band.  See \S2 of \cite{Cha09a} and  \S2 of \cite{Sae09} for details on
source and background extraction regions used for measuring $N_{sc}$.}

\tablenotetext{c}{The fluxes (in units of $10^{-13}$~ergs cm$^{-2}$
s$^{-1}$) in the 0.5--8~keV \OF\ band are obtained using the
best-fit absorbed power law model with two absorption lines. 
The details of these fits are presented in Table~3 of \cite{Sae09} and in Table~2 of \cite{Cha09a}. The fluxes  have been corrected for  Galactic absorption.}

\end{deluxetable*}


Gravitational lensing of \apm\ produces three images with the two brightest ones separated by $\sim$ 0.5\arcsec\ (see Figure~\ref{fig:iapm}). The time-delay between images A and B of \apm\ is estimated to be of the order of a few hours \citep[e.g.,][]{Mun01}. The Eddington ratio of \apm\ can be estimated from an assumed value of  $\Gamma \sim 2$ (where $\Gamma$ is the X-ray photon index), which is close to the mean value estimated from observations \citep{Sae09, Cha09a},  and  using the $L_{\rm bol}/L_{\rm Edd}$ vs. $\Gamma$ correlation found in RQ quasars  \citep[e.g.,][]{Wan04,She06,She08}. 
The $L_{\rm bol}/L_{\rm Edd}$ vs. $\Gamma$ correlation for \apm\ implies $L_{\rm bol}/L_{\rm Edd} \sim 0.2$, and therefore $M_{\rm BH} \sim 10^{12}   \mu_L^{-1} M_\odot$.  A second independent method of estimating the Eddington ratio  through the use of the observed width of the \CIV\ line indicates a black-hole mass of $M_{\rm BH} \sim 10^{11}   \mu_L^{-1} M_\odot$  \citep[see e.g.,][]{Rie09}. The Eddington ratio derived using the second method is  $L_{\rm bol}/L_{\rm Edd} \sim 2$, implying the source is  accreting at a super-Eddington rate.  However caution must be taken with this result because the estimates of $M_{\rm BH}$ based on \CIV\ emission-line widths is subject to a lot of systematic uncertainties \citep[i.e., it contains non-virial components such as the disk wind that can contribute to the broadening, e.g.,][]{She08}.  There are also conflicting reported estimates of the magnification parameter ($\mu_L$).  Optical observations indicate $\mu_L \sim 100$ \citep[e.g,][]{Ega00}, while observations of CO emission yield  $\mu_L\sim 4$  \citep[e.g.,][]{Rie09}.  
In this work we try, as far as possible, to present our results as a function of the magnification 
factor, however when this is not possible a value of $\mu_L=100$  is assumed \citep[as in,][]{Ega00}. The absorption corrected optical to  X-ray power-law slope\footnote{\aox\ is defined as the slope of a hypothetical power law extending between 2500\AA\ and 2 keV in the AGN rest frame, i.e. $ \aox ={\rm log} \frac{F_\nu(2keV)}{F_\nu (2500 \Am)}/{\rm log} \frac{\nu(2keV)}{\nu (2500 \Am)}=0.3838~{\rm log} \frac{F_\nu(2keV)}{F_\nu (2500 \Am)} $}, of \apm\  is \footnote{The  $\aox-l_{\rm UV}$ relation described in \citep{Str05, Ste06} predicts that $\aox=(-0.137 \pm 0.008)~{\rm log}~l_{2500 \Am} +(2.638\pm 0.240)$. Using \cite{Irw98} work we estimate that ${\rm log}~l_{2500\Am} \sim 31.7$ ($\mu_L \sim 100$) and therefore we obtain $\aox=-1.7 \pm 0.3$ in accordance with \cite{Dai04} value.} $\aox \sim -1.8$  \citep[e.g.,][]{Dai04}. The relatively soft SED of \apm\ combined with its nearly large Eddington ratio provide ideal conditions for the production of radiatively driven winds \citep[see e.g.,][and \S \ref{S:SEDs}]{Eve05}.

\subsection{Results from eight deep 
X-ray observations of  \apm} \label{S:Xobs}

This section summarizes results from 2 \chandra, 3 \xmm\ and 3 \suzaku\ observations of \apm\ (8 in total; see Table~\ref{tab:obid} for details). The data reduction of the \chandra\ and \xmm\ observations 
is described in \cite{Cha09a}  and the reduction of the \suzaku\  observations in \cite{Sae09}.
  Note that from here on when we refer to a particular observation we will use the same names defined in \cite{Sae09} and \cite{Cha09a}. Therefore the three \suzaku\ observations are referred to as OBS1, OBS2 and OBS3, the three \xmm\ observations are referred to as Epoch~2, Epoch~3 and Epoch~4 and the two \chandra\ observations are referred to as Epoch~1 and Epoch~5 (see Table~\ref{tab:obid} for details). 

\begin{deluxetable*}{cccccccc}  \tabletypesize{\small}
\tablecolumns{12} \tablewidth{0pt} \tablecaption{
The minimum and maximum
energies and velocities of the high-energy absorption features in \apm$^{a}$.
\label{tab:vmnx}}

\tablehead{
 \colhead{OBS} & \colhead{Instrument} & & \colhead{$E_{\rm min}$} & \colhead{$E_{\rm max}$} & \colhead{$v_{\rm min}$} & \colhead{$v_{\rm max}$}  & $\Gamma$ \\
 & & & [keV] & [keV] & [$c$] & [$c$]
}

\startdata

OBS1  & XIS FI & &7.52$\pm$0.36 & 12.16$\pm$0.52 & 0.11$\pm$0.05 & 0.53$\pm$0.03  & 1.94$\pm$0.04  \\
OBS2 & XIS FI & & 7.88$\pm$0.48 & 11.57$\pm$0.51 & 0.16$\pm$0.06 & 0.50$\pm$0.03 & 2.02$\pm$0.03 \\
OBS3 & XIS FI & & 7.51$\pm$0.32 & 11.70$\pm$1.06 & 0.11$\pm$0.04 & 0.51$\pm$0.07 & 1.94$\pm$0.05 \\

OBS1  & XIS BI & &   7.24$\pm$0.51 & 12.47$\pm$1.12 & 0.08$\pm$0.07 & 0.55$\pm$0.06 & 1.95$\pm$0.07\\
OBS2 & XIS BI  & & 8.74$\pm$0.58 & 12.24$\pm$1.16 & 0.26$\pm$0.06 & 0.54$\pm$0.07 & 1.91$\pm$0.06\\
OBS3 & XIS BI   & & 7.53$\pm$0.82 & 12.40$\pm$0.92 & 0.12$\pm$0.11 & 0.54$\pm$0.05 & 1.92$\pm$0.05\\

Epoch1 & ACIS S3 & & 8.05$\pm$0.11 & 10.64$\pm$0.19 & 0.18$\pm$0.01 & 0.43$\pm$0.01 &  1.74$\pm$0.03\\\
Epoch2  & EPIC pn & &7.54$\pm$0.42  & 15.04$\pm$0.97 & 0.12$\pm$0.06 & 0.67$\pm$0.04  &  1.89$\pm$0.03 \\
Epoch3 & EPIC pn & & 6.42$\pm$0.48  & 17.94$\pm$1.81 &  $<0.03$ & 0.76$\pm$0.04   &  2.03$\pm$0.04 \\
Epoch4 & EPIC pn & & 6.94$\pm$0.20  & 16.31$\pm$0.90 & 0.04$\pm$0.03 & 0.71$\pm$0.03 &  2.11$\pm$0.02\\
Epoch5 & ACIS S3  & & 7.36$\pm$0.30 & 15.35$\pm$1.13 & 0.09$\pm$0.04 & 0.68$\pm$0.04 &  1.94$\pm$0.04\\
\enddata
\tablenotetext{a}{The model used to estimate $E_{\rm min}$ and $E_{\rm
max}$ is a power-law with Galactic absorption, intrinsic absorption, and two absorption lines (APL+2AL). The fits of this model are presented in Table~3 of \cite{Sae09} and in Table~2 of \cite{Cha09a}.}
\end{deluxetable*}

 
Our analysis of the X-ray observations of the BAL quasar APM 08279+5255 indicates strong and broad absorption at rest-frame energies of 1--4~keV (low-energy) and 7--18~keV (high-energy).  The medium producing the low-energy absorption is a nearly neutral absorber with a column density of \hbox{$\lnhcm \sim 23$}. However, since \apm\ is at a high redshift of \hbox{$z=3.91$} it is difficult to constrain the ionization state of the low-energy absorber.   The absorption signatures of this complex low energy absorber are shifted outside the range of the observed X-ray band (typically between 0.3--10~keV).  The high-energy absorption features are easily detected in the 8 deep X-ray observations of \apm. In every case these features satisfy $\rm EW/\sigma_{EW} \gtrsim 3$ \citep[see,][]{Sae09, Cha09a}\footnote{EW is the equivalent width obtained using the best fitted model of the high energy absorption feature. The equivalent width (EW) is defined as $EW=\int \frac{F_c-F_E}{F_c}dE$, where $F_c$ is the continuum flux and $F_E$ is the flux in the absorber.},  and therefore they satisfy realistic limits for significant detections \citep{Vau08}.

Chartas el al. 2002 have interpreted the high-energy X-ray BALs as being produced by absorption of highly ionized iron such as 
 \FeXXV\  K$\alpha$ ($1s^{2} -1s2p$;  6.70~keV) and/or \FeXXVI\  ($1s-2p$; 6.97~keV) launched very near an ionizing compact central source.
  Evidence to support this interpretation is the significant variability in the strength and energy of the X-ray BALs over short time-scales present in many of the observations. 
This variability is found to be as short as  $\sim$~3~days \RF\  in the \chandra\ and \xmm\ observations,  and  $\sim$~month in the \suzaku\ observations.  This fast variability presents a strong argument in favor of a wind that originates 
from a distance of a few times the Schwarzschild radius\footnote{The time-scale of the flux variability provides an indication of the outflow launching radius  from the central engine \citep[see e.g.,][]{Sae09}. The size of this region should be approximately the time scale of the variability times the speed of light. Variability over a period of a $\sim$week implies a launching radius $\sim$~$6 R_S$. In the last expression we estimated the Schwarzschild  radius ($R_S$) of \apm\ using $M_{\rm BH} \sim 10^{10} M_\odot$ (obtained assuming $\mu_L \sim 100$; see  \S \ref{S:gapm}). Possible flux variability caused by the time-delay between the images is shorter than $R_S/c$.}.

In the \suzaku\ observations the variability of the high energy absorber is present at energies close to 7~keV \RF.  On the other hand the \chandra\ and \xmm\ observations  show variability of the high-energy absorption profiles at both:  low energies (i.e.  \RF\ energies  $\sim 7$~keV) and high energies (i.e. \RF\ energies $\gtrsim 9$~keV).
A time-variable outflow provides a plausible explanation for the changes in shape on the absorption features in past X-ray observations of \apm\ \citep{Cha02,  Has02}. An even stronger case is presented in \cite{Cha09a} and \cite{Sae09}, where it can be seen clearly that in some cases the appearance of the high-energy absorption feature can take the shape of a notch, an edge or two absorption lines.  Our spectral analysis of the 8 deep observations also indicates variability of the parameters defining the high energy attenuation which is likely due to a change in the outflow velocity of the absorber \citep{Sae09, Cha09a}. The short time-scale  ($\sim$week in the rest-frame)  of the variability combined with the high ionization of the absorbing material which is moving at relativistic speeds imply that the absorbers are launched from distances $\lesssim 10~R_S$ from the central source \citep[see e.g \S4 of][]{Sae09}.  The short time-scale of this variability in the \suzaku\ \citep{Sae09} and \xmm\ \citep{Cha09a} observations also indicates that this absorber should be strongly accelerated.  
The global covering factor of these winds should be low $\lesssim 20 \%$ based on the absence of emission features from the outflowing ionized gas \citep[see e.g.][]{Cha09a}.

The minimum and maximum projected velocities ($v_{\rm min}, v_{\rm max}$)
of the outflow are estimated from the minimum and maximum
energy ranges ($E_{\rm min}, E_{\rm max}$) of the high-energy absorption
features in \apm. We obtained $E_{\rm min}$ and $E_{\rm max}$
from our spectral fits assuming first the two absorption-line
(APL+2AL) model.  Specifically, based on the best-fit values of an absorbed \PL\ model with two absorption lines (from Table~3 of Saez, Chartas \& Brandt 2009 and Table~2 of Chartas et al. 2009), we obtain $E_{\rm min} = E_{\rm abs1} - 2\sigma_{\rm abs1}$ and $E_{\rm max} = E_{\rm abs2} + 2 \sigma_{\rm abs2}$. As in \cite{Cha09a},  we
estimate the line of sight projected velocities $v_{\rm min}$ and $v_{\rm max}$ assuming the absorption arises from highly blueshifted Fe~xxv~K$\alpha$ ($E_{\rm lab} = 6.7$~keV). 
In Table~\ref{tab:vmnx} we show the values obtained for $E_{\rm min}$, $E_{\rm max}$, $v_{\rm min}$ and $v_{\rm max}$. In this table we also add a column  with the fitted values of $\Gamma$ (based on  the APL+2AL model).  
Using $E_{\rm max}$ $\sim$ 18~keV (the maximum observed value), we constrain the maximum angle\footnote{The Doppler-shift formula
 predicts that given a fixed ratio of
$E_{\rm lab}$/$E_{\rm obs}$$\equiv$$R_{\rm lo}$ (where $E_{\rm lab}$ and $E_{\rm obs}$  is the energy of the absorption line in the \RF\ and \OF\ respectively) the maximum angle
between our line of sight and the wind direction is given by
$\theta_{\rm max}={\rm cos}^{-1}(\sqrt{1-R_{lo}^2})$.} 
between our line of sight and the wind direction to be less than $22^{\circ}$.

As we indicated in \cite{Cha09a}  (and from Table~\ref{tab:vmnx}) there is a hint that changes in the photon index ($\Gamma$) may be positively correlated with the changes of the maximum velocity of the outflow ($v_{\rm max}$). This possible trend between $\Gamma$ versus $v_{\rm max}$ is shown in Figure~10 of \cite{Cha09a}.
In \S \ref{S:nro} we recalculate the velocities found in Table~\ref{tab:vmnx} with a model based on \cloudy\ simulations of a near-relativistic outflow. 
In \S \ref{S:SEDs} we provide an interpretation of the possible trend between maximum outflow velocity and $\Gamma$.

\section{Photoionization models of near-relativistic outflows}  \label{S:nro}

\subsection{Motivation}

From the  X-ray analysis of our 8 observations of  \apm\ \citep[see,] [for more details]{Sae09, Cha09a}, and past observations of this object and \pg\ \citep[e.g.,][]{Cha07a}, it became clear that a more sophisticated spectral model was needed to fit the X-ray BALs. 
Currently there are no proper tools available in the spectral fitting package \xspec\ to model absorption profiles resulting from near-relativistic outflows. In \cite{Sae09} we used photoionization models created from the photoionization code \xstar\ 
\citep[see, e.g.,][]{Kal01} to fit the  broad absorption feature found between 7$-$18~keV rest-frame in APM 08279. This spectral analysis did not include relativistic  corrections, and does not contain a realistic model of the outflow. In this section we describe new software code that we 
have developed that provides a more realistic description of radially accelerated near-relativistic outflows.
Our main goal is to better  constrain important parameters describing near-relativistic outflows; among these are, the velocity profile, the ionization parameter and the column density of the observed X-ray wind.

\subsection{Description of the code}

Our quasar outflow code is based on a  multilayer approach which mimics the absorption and scattering through a near-relativistic  outflow using the photoionization code \cloudy\ \citep{Fer98}.
An existing quasar outflow code called {\sc xscort} developed by Schurch \& Done 2007 follows a similar approach and
is based on the photoionization code \xstar.

In our quasar outflow code we use \cloudy\ simulations to approximate the absorption signature produced by an outflowing medium that is radiatively accelerated from a central source.  We describe the outflowing medium 
with a set of absorption layers with specific velocity profile and ionization states. The details of the approximations and assumptions used in our quasar outflow model are described in the following paragraphs.

\subsubsection{The case of one absorption layer}

We approximate the attenuation through the outflowing ionized gas
as the absorption through a series of layers of different velocities and ionization states. 
We begin by calculating the output absorption profile assuming the wind is made up of a single layer and
then generalize our model by dividing the outflowing wind into multiple layers. We assume the
layer is at a distance $R$ from the continuum source, has a thickness
$\Delta R \ll R$, a density $n$, and is moving at a velocity $v$ away from
it. We further assume that spectral energy distribution of the source is a power-law 
($L_\nu \propto \nu^{\alpha}$) with spectral index $\alpha=-1.0$. 
From here on, unless mentioned otherwise, this is the SED that we will assume in our simulations.
The ionization parameter \citep[][see Appendix~\ref{ap:ionp}]{Tar69} measured in the absorption layer's rest-frame is

\begin{equation} \label{eq:IP}
\xi=\frac{4 \pi F'_I}{n'}
\end{equation}

where primed quantities  $n'$ ($n'= \gamma n$) and $F'_{I}$ refer to the density and the incident ionizing flux in the 
layer's rest-frame, respectively. Non primed parameters are assumed to be in the \RF\ of the luminous source. 
Using the fact that $I_\nu/\nu^3$ is a Lorentz invariant, the incident ionization flux in a layer's rest-frame  is given by\footnote{Since $F_\nu \propto I_ \nu$, $\nu/\nu'=D$; where $D=\gamma^{-1}(1-v/c)^{-1}$ and  $I_ \nu=I_{\nu'} \nu^3/\nu'^3=I_{\nu'}D^3$. Therefore $F_I'=\int_{\nu_0'}^{\nu_1'}F_{\nu'}d\nu'=D^{-4} \int_{\nu_0}^{\nu_1}F_\nu d\nu$, where $\nu_0'=1$~Ry and  $\nu_1'=1000$~Ry; $\nu_{0,1}=D \nu'_{0,1}$.  Conversely, to calculate the incident flux in the layer's rest-frame,  we integrate the flux in the luminous source \RF, and multiply by a factor of $D^{-4}$. Notice that if the flux is a power law, i.e. $F_\nu=a \nu^{\alpha}$,  then $F_I'=aD^{-4}{\rm ln}(\nu_1'/\nu_0')$ if $\alpha=-1$, and $F_I'=aD^{\alpha-3}(\nu_1'^{\alpha+1}-\nu_0'^{\alpha+1})/(\alpha+1)$ if $\alpha \neq -1$. Therefore in the case that the SED is a power-law, 
 the ionizing flux in the  rest-frame of the absorbing layer can be found by integrating the flux in the luminous source \RF\ between 1~Ry and 1000~Ry, and then multiply by $D^{\alpha-3}$ where $\alpha$ is the spectral index. }

\begin{equation}
F'_I=  \gamma^4(1-v/c)^4 F_I.
\end{equation}

The column density (in any reference frame\footnote{$N_{\rm H}=n\Delta r=(n{\gamma}) (\Delta r/\gamma)=n' \Delta r'$}) of the layer is given by $N_{\rm H}=n\Delta r$. The incident flux on a layer is calculated between 1--1000 Ry in the 
rest-frame of the moving layer.\footnote{The Rydberg is a unit of energy defined in terms of the ground-state energy of an electron for the hydrogen atom, $1~{\rm Ry} \approx 13.6 eV$.} 
From the incident flux in the rest-frame of the layer and from the use of equation (\ref{eq:IP})  we calculate the ionization state of this layer. 
We next input the derived ionization state and column density of the layer into the photoionization code \cloudy\  to obtain the output spectrum  of the layer in its rest-frame and transform it to the rest-frame of the luminous source.

\subsubsection{A model to describe X-ray absorption profiles.}

The X-ray spectra that we have obtained from observations of several BAL quasars do not show any emission lines associated with the X-ray BALs  \citep{Cha09a}. This is expected since BAL quasar outflows in general  have relatively small global covering factors \citep{Hew03}. Therefore, we will concentrate on absorption-dominated outflows neglecting emission features in our model. 
For our \cloudy\ code, we will assume an approximately constant ionization parameter in the outflow. 

We assume that the region of the outflow that contributes to the  observed \FeXXV\ and/or \FeXXVI\ absorption is localized in a  relatively thin multilayer of thickness $\Delta R$ at a distance $R$ from the black hole.
The justification behind the assumption that $\Delta R \ll R$ in our model is that the presence of  \FeXXV\  is likely to occur in a relatively narrow region away from the source where the ionization parameter and density are such that  \FeXXV\  is abundant at these distances.
Even though the ionization parameter is expected to fall off with distance along the flow as $1/R^{2}$ it remains relatively constant within the multilayer since $\Delta R \ll R$.\footnote{In our simulations we assume that the ionization luminosity is $L_I=10^{44}~\lumin$, therefore for a layer with $\lip=3.0$, $n = 10^{12}~\cc$ (initial density assumed in our runs, see \S \ref{S:tria}),  and $\lnhcm =23$, we will typically have that the ratio of thickness to distance of the layer is $\Delta R/R \approx 0.001 $ and consequently $\Delta \xi \approx -2$. This means that the ionization parameter with typical value of $\lip \sim 3$ varies through the outflow $\sim 0.001$~dex due to the radial dependency.}
To compensate for changes in flux across the multilayer due to relativistic effects (e.g., beaming effects) we  adjust the density $n_{i}$ of each layer such that the ionization parameter remains approximately constant within the multilayer  $\Delta R$.  In the case that the region of the outflow that we are modeling is not physically thin, our approach will still be valid as long as the ionization parameter along this region does not change appreciably.

From equation~(\ref{eq:IP}) the ionization parameter of each layer is $\xi_i=4\pi F_{\rm I,i}'/n'_i$; where $F_{I,i}'$ is the incident flux in the rest-frame of layer $i$ estimated from the flux coming from layer $i-1$ after taking into account relativistic effects.  
The density of each layer, is determined from an a priori defined velocity profile $v(r)$.  In our multilayer approach, the density of each layer is $n_i \propto \gamma_i^{3} (1-v_i/c)^{4}$ for $i=0,1..(N_l-1)$, were $N_l$ is the number of layers.  We note that the assumed density profile  adjusts the ionization parameter so that it does not vary across the multilayer due to relativistic effects.
However, the ionization parameter will still decrease between $0.02-0.40$~dex  mainly due to attenuation (see \S \ref{S:lays}). 
For our case we concentrate on simulated outflows with ${\rm log~} \xi \gtrsim 3$. As long as each layer\footnote{We use $\approx$~100 layers to generate the absorption profile of a simulated wind, therefore each layer has a column density of ${\rm log}~\Delta \nh_i \lesssim 22$.}  used in our calculations has  a column density of  ${\rm log}~(\Delta \nh_i/{\rm cm}^{-2}) \lesssim 23$ it can be assumed to be almost transparent (or thin) to radiation (see Appendix~\ref{ap:nhma}).  As we describe in \S \ref{S:lays} the assumption of optically thin layers will allow us to more greatly reduce the time needed to compute the absorption signature of the multilayered outflow. \\
In this work we assume that the outflow velocity profile will have a $p$-type form given by

\begin{equation} \label{eq:vNH}
v(\hat{N}_{\rm H})=v_{\rm min}+(v_{\rm max}-v_{\rm min})(\hat{N}_{\rm H}/\nh)^p,
\end{equation}

where $\nh$ is the total column density of the simulated wind ($0 < \hat{N}_{\rm H} < \nh$). 
We note that given the assumptions described here, and the moderate S/N of the absorption profile in \XR, we do not expect to obtain significant constraints on the 
acceleration mechanism of the outflow by using this profile.

\subsubsection{Passing a continuum spectrum from one absorbing layer to the next} \label{S:lays}
 
When passing the continuum spectrum from one layer ($i$-1) to the next ($i$) we first transform the source \RF\ spectrum coming from layer $i$-1 to the rest-frame of layer $i$. 
In the rest-frame of  layer $i$ we remove absorption and add emission 
(in cases where we do consider emission) to the incident spectrum given the column density and ionization parameter of the layer.
The column density  of layer $i$ is obtained from the velocity profile, and its ionization parameter is calculated from the input spectrum in its \RF. Finally we transform the output spectrum coming from layer $i$ back to the source \RF.

Since the density of each layer is chosen to correct the attenuation of the flux due to relativistic effects, the ionization parameter will mostly decrease due to the absorption of the outflow across the layer. The decrease of the ionization through the multilayered outflow due to absorption will be greater in winds that are less ionized and with higher column densities. For example, a wind with $\lnhcm=22.75$ with initial ionization parameter of $\lip_i=3.25$ will have a final ionization parameter of $\lip_f \sim 3.21$, however for  $\lnhcm=22.75$ and $\lip_i = 3.75$ the final ionization parameter is $\lip_f \sim 3.73$. Additionally a wind with $\lnhcm=23.75$ and $\lip_i=3.25$ will have $\lip_f \sim 2.84$, however if  $\lnhcm=23.75$ and $\lip_i=3.75$ then $\lip_f \sim 3.53$.

After correcting the spectrum incident on layer $i$ for the effects of relativistic beaming, we estimate the absorption and emission from  layer $i$ using simulations performed with the photoionization code \cloudy. These \cloudy\ simulations are implemented using the unabsorbed source spectrum (power-law with $\alpha=-1$ in our case) and 
various values of the ionization parameter. 
The main advantage of this approximation is a dramatic shortening in the time it takes to obtain a profile. This speed-up is possible because the photoionization runs at each layer 
use a library of preexisting \cloudy\ runs (see next paragraph for details).  
If the flux emitted by the source is a power-law, the flux received by a moving layer is also a 
power-law with the same spectral index as the source
as long as the SED is not significantly attenuated in an energy dependent way. 
In addition, if each layer of the multilayered media is approximately optically thin then the state of the gas depends primarily on the ionization
parameter and secondarily on the SED \citep{Tar69}. This dependence indicates that our approach is at first order a good approximation as long as the SED along the multilayered outflow does not differ appreciably from the incident SED at the first layer. However, for low ionization states ($\lip < 3$) and large total column densities ($\lnhcm \gtrsim 23.5$), the SED along the multilayered outflow will deviate appreciably from the initial incident SED, and therefore,  our approximation breaks down.

In order to avoid invoking \cloudy\ too many times ($\sim100$) to simulate the signature of the multilayered absorber, we created a library of spectra  to speed-up our calculations. This library contains a set of simulated transmitted spectra through absorbers having a range of ionization parameters between $1.8-4.1$~dex (\lip\ steps of 0.01~dex), column density of $\lnhcm=19$,  density of $n_{\rm H}=10^{12}$~\cc\ and an input SED that has the form of a power law ($L_\nu \propto \nu^{\alpha}$) in the energy range between $1-10^4$ Rys with spectral index $\alpha=-1$. The value of the density of the absorber  ($n_{\rm H}=10^{12}$~\cc) is chosen to be close to the value of the density 
predicted by 2-d numerical simulations of radiative winds by \cite{Pro00, Pro04}. 
The simulations used to construct this library (from here on referred to as library runs), as well as all the 
simulated profiles described in the next sections, are calculated using standard solar metallicities.
We primarily used \cloudy\ to perform the photoionization calculations but we also compared our results with those generated by the photoionization code \xstar. 

In order to reduce the number of realizations in our library, we performed these runs assuming a specific turbulent velocity of $v_{\rm turb}=1500~\kms$.  
Our choice of a specific turbulent velocity requires that the relative speed $\Delta v_i'$ between layers $i$ and $i+1$  to be the same for every layer in the simulations of multilayered outflows\footnote{The relative velocities of two adjacent layers according to special relativity is $\Delta v_i'=(v_{i+1}-v_i)/(1-v_i v_{i+1}/c^2)$.} (i.e. $\Delta v_i'= \Delta v'$; see Appendix~\ref{ap:Dvco}).
In our simulations we use a finite number of layers to describe the outflow,  each layer having a slightly different velocity following the adopted velocity profile. The distribution of the velocity differences between adjacent layers is assumed to be uniform and to extend between 0 and a maximum velocity difference of  $\Delta v'$. The  velocity difference of  $\Delta v'$ is kept lower than $v_{\rm turb} \sqrt{12}$; where $v_{\rm turb}  =1500~\kms$ is the  assumed turbulent velocity \citep[see][]{Sch07}.
By selecting a turbulent velocity we set a lower limit on the number of layers used in simulating a BAL spectrum.\footnote{For example if the velocity profile has $v_{\rm min}=0$ and $v_{\rm max}=0.7 c$ we require that the number of layers \hbox{$N_l > (v_{\rm max}- v_{\rm min})/(\sqrt 12 v_{\rm turb})$}, i.e.  with $v_{\rm turb}=1500~\kms$,  the minimum number of layers that we can use for this case is forty. } The simulations making up our library are all performed using the same column density ($\lnhcm=19$).
Consequently, the opacity of a layer $i$ (i.e., the layer's rest-frame absorption) is calculated by first obtaining the opacity from the library at the ionization state of the layer, and second, by scaling the opacity obtained (from the library)  to the column density of the layer\footnote{For thin layers, the opacity of a layer ($\tau_{2}$) can be obtained by scaling the opacity of a different layer ($\tau_{1}$) by the ratio of the column densities of the two layers (i.e. $\tau_2=\tau_1 \nh_2/\nh_1$).}. 
This approximation is justified given that in the range of densities of our simulations ($8 \lesssim \lden \lesssim 12$) the dependency of the opacity on $n_H$ is negligible, and therefore, the opacity of a layer $i$ can be found from the library just by knowing its ionization state and column density.  Emission spectra
 may also be obtained from our library runs, however for our study we mostly concentrated on the absorption spectra. 
 We focus on the absorption spectra since the main goal of our study is to fit the spectral signatures produced by winds 
in which emission lines are relatively weak implying low global covering factors.
Additionally, we assume that the line-of-sight covering factor is equal to one.

\subsubsection{Testing our relativistic outflow code} \label{S:tria}

We first  generate the absorption profiles for  a linearly accelerated outflow with $v_{\rm min}=0$, $v_{\rm max}=0.3c$ ($v \propto \nh$), $\nh=3 \times 10^{23}$~\cmsq ,  initial density $n_0=10^{12}$~\cc, and  values of $\lip=$ 2.75, 3.00, 3.25, 3.50, 3.75 and 4.00.  These profiles have been generated to compare the results of our code with those presented in \cite{Sch07}. In order to make a fair comparison with the results of \cite{Sch07},  we use a library of \xstar\ runs using a power-law model with $\alpha=-1.4$. This library of \xstar\ runs, has the same characteristics as the \cloudy\ library described in \S \ref{S:lays}.  Our simulated X-ray absorption profiles are very similar to the ones computed  by Schurch \& Done 2007  under the assumption of a multilayer outflow with spherical symmetry (see Appendix~\ref{ap:Schu}). 

\begin{figure}
 \includegraphics[width=8.6cm]{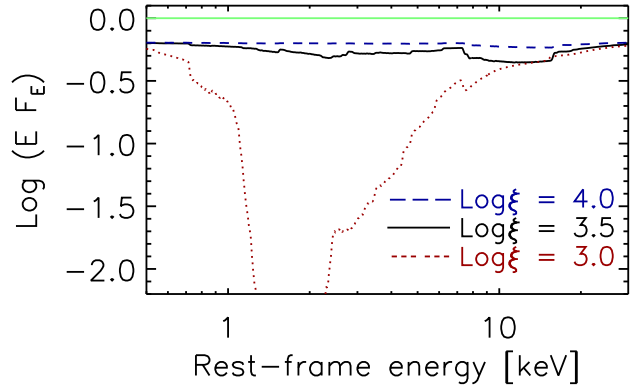}
        \centering
    \caption[X-ray spectra derived from multilayered \cloudy\ simulations
of near-relativistic outflows.]{X-ray spectra derived from multilayered \cloudy\ simulations
of near-relativistic outflows. The assumed input spectrum is a power-law
with $\Gamma=2.0$. The outflow has been accelerated between
$0.1c-0.7c$ using a $p$-type velocity profile with $p=1.0$. The total column
density of the outflow is $\lnhcm = 23.75$. We have calculated
the output spectra for 3 different values of the ionization
parameter, $\lip$ = 3.0, 3.5 and 4.0.}
      \label{fig:run1}
\end{figure}

\begin{figure}
\includegraphics[width=8.6cm]{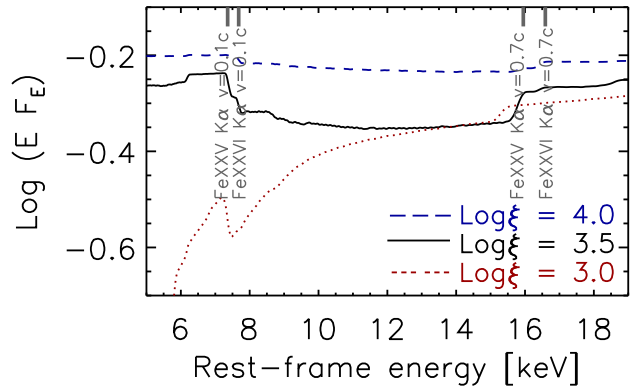}
        \centering
    \caption[X-ray spectra derived from multilayered \cloudy\ simulations
of near-relativistic outflows (zoom in).]{X-ray spectra derived from multilayered \cloudy\ simulations
of near-relativistic outflows. This is a zooming of  Figure~\ref{fig:run1} in the region where we find iron absorption. The assumed input spectrum is a power-law
with $\Gamma=2.0$. The outflow has been accelerated between
$0.1c-0.7c$ using a $p$-type velocity profile with $p$=1.0. The total column
density of the outflow is $\lnhcm = 23.75$. We have calculated
the output spectra for 3 different values of the ionization
parameter, $\lip$ = 3.0, 3.5 and 4.0.  For reference the observed blueshifted energies of the \FeXXV\ K$\alpha$ and \FeXXVI\ K$\alpha$  lines with $v = 0.1c$ and of the \FeXXV\ K$\alpha$ and \FeXXVI\ K$\alpha$ lines with $v = 0.7c$ are highlighted.} 
      \label{fig:zoom}
\end{figure}

\begin{figure}
\includegraphics[width=5.0cm]{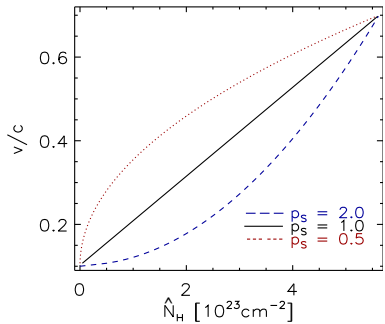}
        \centering
    \caption[Outflow velocity as a function of column density through the outflow for 3 different values of the exponent of the $p$-type velocity profile, $p$ = 0.5 (dotted line), 1.0 (solid line) and 2.0 (dashed line).]{Outflow velocity as a function of column density through the outflow for 3 different values of the the exponent of the $p$-type velocity profile (see equation \ref{eq:vNH}), $p$ = 0.5 (dotted line), 1.0 (solid line) and 2.0 (dashed line).  
    We assumed the outflow has a  launch velocity of 0.1~$c$, a terminal velocity of 0.7~$c$ and  a total column density of $\lnhcm=23.75$. }
      \label{fig:ppro}
\end{figure}

\begin{figure}
\includegraphics[width=8.4cm]{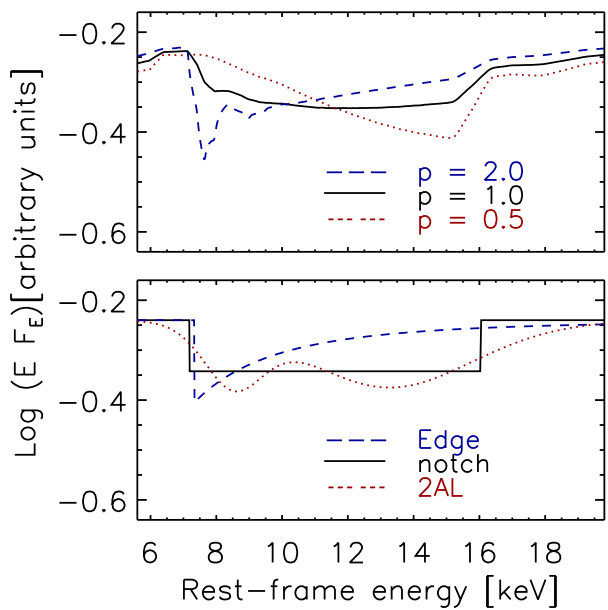}
        \centering
    \caption[{Upper panel:}  X-ray spectra produced from multilayered \cloudy\ simulations of near-relativistic outflows.  {Lower panel:} X-ray spectra resulting from the best fitted models used in \cite{Cha09a}  to fit the high energy absorption of \apm\  in Epoch~3. ]{{Upper panel:}  X-ray spectra produced from multilayered \cloudy\ simulations of near-relativistic outflows.
    The assumed input spectrum is a power-law
with $\Gamma=2.0$ (solid line). The outflow has been accelerated between
0.1$c$$-$0.7$c$ using an exponent of the velocity profile with $p$=1.0. The total column
density of the outflow is $\lnhcm = 23.75$. We have calculated
the output spectra for 3 different values of the exponent $p$, $p$ = 0.5 (dotted line), 1.0 (solid line) and 2.0 (dashed line). 
     {Lower panel:} X-ray spectra resulting from the best fitted models used in \cite{Cha09a}  to fit the high energy absorption of \apm\  in Epoch~3. The solid line is for the notch model, the dotted line corresponds to the two absorption line model and the dashed line is for the absorption edge model.}
      \label{fig:run2}
\end{figure}

We generate a second run of simulations based on our results presented in \S \ref{S:Xobs}. For this purpose,   we simulate absorption profiles  with $v_{\rm min}=0.1 c$, $v_{\rm max}=0.7 c$, $n_0=10^{12}~\cc$, $\lnhcm = 23.75$, and three values of the ionization parameter $\lip=3.0, 3.5, 4.0$.  The outflow parameters have been chosen approximately in accordance to the estimated values based on one of the \xmm\ observations of \apm\  \citep[Epoch~3 from][]{Cha09a}.
To generate \hbox{notch-type} absorption profiles we chose $p=1$ for these  simulations. The resulting profiles are presented in Figure~\ref{fig:run1}.  In Figure~\ref{fig:zoom} we zoom in on the iron blend feature which shows up at energies between 7.5$-$16.5~keV.  The  resonance absorption lines  responsible for the features are mainly  \FeXXV\  K$\alpha$ ($1s^{2} -1s2p$;  6.70~keV) and \FeXXVI\  K$\alpha$ ($1s-2p$; 6.97~keV).
From Figures~\ref{fig:run1} and \ref{fig:zoom} it is clear that the 
strength of the iron blend feature is larger for $\lip = 3.5$. The strength of this feature is large for ionization parameters in the 
range of $3 \lesssim \lip \lesssim 4$, since at lower ionization levels the abundance of highly ionized iron atoms (e.g.,  \FeXXV\ and \FeXXVI ) is too small to produce any appreciable feature in the spectra.  At $\lip \gtrsim 4$ the atoms in the gas become stripped of almost all their electrons, and therefore, the resulting spectra contain very few \hbox{bound-bound} and \hbox{bound-free} absorption features. When $\lip \lesssim 3$ there is substantial absorption present at \RF\ energies $\lesssim$~2~keV. As a consequence, in a heavily absorbed source like \apm, the low-energy  ($\lesssim$~2~keV) absorbed component of the spectrum, possibly due to non-outflowing material, will overlap with the absorbed component produced by the wind \citep[e.g.,][]{Sae09, Cha09a}. However, the high-energy absorption iron trough (above $\sim 7$~keV)   is not contaminated by absorption lines from non-outflowing material, and consequently, it provides a clean signature of the outflowing material.

The absorption profiles described in \S \ref{S:Xobs} contain a diversity of shapes, sometimes resembling notches, edges or two absorption lines. Given the variety  of shapes of the absorption profiles we attempted to emulate these different shapes by generating simulations of outflows  with $v_{\rm min}=0.1 c$ and $v_{\rm max}=0.7 c$, $\lnhcm=23.75$, $\lip = 3.5$ and three values of $p$, $p=0.5,1$ and 2. The output spectra of the resulting velocity profiles (see Figure~\ref{fig:ppro} ) are presented in Figure~\ref{fig:run2}. In the lower panel of  Figure~\ref{fig:run2} we also show as a reference three model spectra used to fit  Epoch~3. The three different models used are a notch, an edge and a two absorption line model \citep[see Table 2 from][]{Cha09a}.

\subsection{Fits to the observed spectra of \apm\ using our quasar outflow model}

In this section we present results from fits to spectra obtained from {\it XMM-Newton}, {\it Chandra} and {\it Suzaku} observations of \apm\ using our thin multilayered outflow model. 
We created \xspec\ table models\footnote{A table model consist of a file that contains an $N$-dimensional grid of model spectra with each point on the grid having been calculated for particular values of the $N$ parameters in the model.  \xspec\ will interpolate on the grid to get the spectrum for the parameter values required at that point in the fit.}  from a large number of simulated quasar absorption 
spectra produced from running our multilayered model over a large range of 
parameter space of the outflow parameters. Specifically, the parameter space of our simulations covered five important properties of the multilayered outflow model:   
minimum velocity of the outflowing gas $v_{\rm min}$,  maximum velocity $v_{\rm max}$, 
ionization parameter $\xi$, total column density \nh\ and the value $p$ which defines the
outflow velocity profile (see equation \ref{eq:vNH}).
For our simulations we assume (unless mentioned otherwise) an initial density of the outflowing gas in the multilayer of $n_0=10^{12}$~\cc. 
We note that our results are not sensitive to the assumed initial density of the outflowing gas.  A change of the density by up to two orders of magnitudes from our assumed value 
does not result  in any significant variation of our results.

\subsubsection{\xspec\ table models for quasar outflows}

In order to fit the variety of absorption spectra of \apm\ 
we generated three different types of \xspec\ table models that mainly differ in
the parameter space of the outflow properties that are covered.

The first table model ({\sc windfull.tab}) assumes the presence of only one outflowing 
component. We define as component $i$ of an outflow a multilayer absorber 
at a distance $R_{\rm i}$ from the black hole, with a thickness $\Delta R_{\rm i}$ ($\Delta R_{\rm i} / R_{\rm i} \ll 1$),  a velocity gradient of $\Delta v_{i}$ across it, a column density of $N_{\rm H}$, and an ionization parameter of $\lip_{i}$.
We produced the table model {\sc windfull.tab} by simulating transmitted spectra through a range of outflows with minimum velocities ($v_{\rm min}$) in the range of  0.0$-$0.3$c$ (a simulation for every 0.02$c$ interval), with maximum velocities ($v_{\rm max}$) in the range of 0.3$-$0.9$c$ (step intervals of 0.04$c$),  $\lnhcm$ in the range 22$-$24 (step intervals of 0.25 dex), $\lip$ in the range 2.8$-$4.0 (step intervals of 0.2 dex), and the index $p$ having values of 0.5, 1.0, 2.0 and 5.0.


\begin{deluxetable*}{cccccccccccc} \tabletypesize{\scriptsize}
\tablecolumns{12} \tablewidth{0pt} \tablecaption{
Results from
spectral fits to the three \suzaku~observations of \apm.
\label{tab:mosu}}
\tablehead {
\multicolumn{2}{c}{} & \multicolumn{3}{c}{ FI~~~SPECTRUM$^b$} & & \multicolumn{3}{c}{BI~~~SPECTRUM$^b$} \\
\multicolumn{2}{c}{} & \multicolumn{3}{c}{$ \overline{\hspace{150pt}}$} & & \multicolumn{3}{c}{ $\overline{\hspace{150pt}}$} \\
 \multicolumn{1}{c}{Model$^a$} &  \multicolumn{1}{c}{Parameter} &  \multicolumn{1}{c}{Values OBS1} &  \multicolumn{1}{c}{Values OBS2} & \multicolumn{1}{c}{Values OBS3} & &  \multicolumn{1}{c}{Values OBS1} &  \multicolumn{1}{c}{Values OBS2} &  \multicolumn{1}{c}{Values OBS3}  \\
}

\startdata

MODEL1..   &$\Gamma$  & $1.96_{-0.05}^{+0.04}$  & $2.03_{-0.05}^{+0.06}$  & $1.95_{-0.06}^{+0.06}$ & & $2.03_{-0.15}^{+0.15}$ & $1.93_{-0.05}^{+0.07}$ & $1.91_{-0.07}^{+0.09}$\\
&  ${\rm log}~(N_{\rm H1}/{\rm cm}^{-2})$ &  $22.91_{-0.05}^{+0.09}$  & $22.95_{-0.12}^{+0.12}$  & $22.83_{-0.09}^{+0.11}$ & & $22.76_{-0.13}^{+0.12}$ & $22.79_{-0.09}^{+0.06}$ & $22.78_{-0.09}^{+0.08}$\\
&   ${\rm log}~(N_{\rm H2}/{\rm cm}^{-2})$ &  $23.54_{-0.11}^{+0.10}$  & $23.32_{-0.15}^{+0.18}$ & $23.55_{-0.15}^{+0.12}$ & & $23.82_{-0.16}^{+0.18}$ & $23.42_{-0.15}^{+0.16}$ & $23.68_{-0.22}^{+0.20}$  \\
&  $\lip_2$ & $3.44_{-0.24}^{+0.16}$  & $3.29_{-0.10}^{+0.10}$ & $3.52_{-0.12}^{+0.16}$ & &  $3.59_{-0.17}^{+0.25}$ &  $3.52_{-0.25}^{+0.12}$  &  $3.64_{-0.27}^{+0.22}$\\
&  $v_{\rm min}/c$ & $0.14_{-0.02}^{+0.02}$  & $0.19_{-0.02}^{+0.02}$ &  $0.10_{-0.02}^{+0.02}$ & &  $0.07_{-0.06}^{+0.03}$ &  $0.28_{-0.07}^{+0.04}$ &  $0.13_{-0.06}^{+0.03}$\\
&  $v_{\rm max}/c$ & $0.50_{-0.02}^{+0.04}$  &  $0.51_{-0.04}^{+0.04}$ & $0.49_{-0.08}^{+0.12}$ & & $0.58_{-0.05}^{+0.05}$ & $0.60_{-0.07}^{+0.06}$  & $0.62_{-0.10}^{+0.15}$ \\
& $p$ &   $1.6_{-0.4}^{+0.4}$ & $1.8_{-0.6}^{+0.8}$ & $2.3_{-0.7}^{+1.0}$& &  $1.4_{-0.6}^{+0.6}$  & $2.8_{-1.0}^{+2.2}$ & $2.6_{-1.4}^{+2.0}$\\  
& \lLa  & 46.92$\pm$0.08   & 46.91$\pm$0.07 & 46.93$\pm$0.08 & & 47.10$\pm$0.16 & 46.85$\pm$0.08 & 46.95$\pm$0.10 \\
&   $\chi^2/\nu$ &  60.5/67  & 68.8/60 & 70.2/65 & & 56.1/65 & 89.6/60 & 43.8/65  \\
&   $P(\chi^2/\nu)$ & 0.70 & 0.21 & 0.32 & & 0.78 & 0.01 & 0.98\\
\\

MODEL2..   &$\Gamma$  &  $1.98_{-0.05}^{+0.05}$ & $2.03_{-0.04}^{+0.04}$ & $1.95_{-0.05}^{+0.05}$  & & $2.00_{-0.10}^{+0.10}$ & $1.92_{-0.05}^{+0.05}$ & $1.91_{-0.07}^{+0.07}$\\
&   ${\rm log}~(N_{\rm H1}/{\rm cm}^{-2})$ &  $22.82_{-0.20}^{+0.16}$ &  $22.97_{-0.07}^{+0.07}$ & $22.81_{-0.11}^{+0.13}$  & & $22.76_{-0.15}^{+0.10}$  &  $22.79_{-0.09}^{+0.09}$ &  $22.78_{-0.08}^{+0.06}$ \\
&   ${\rm log}~(N_{\rm H21}/{\rm cm}^{-2})$ &  $23.23_{-0.15}^{+0.15}$  & $23.10_{-0.14}^{+0.20}$ & $23.08_{-0.20}^{+0.18}$ & & $23.50_{-0.18}^{+0.15}$ & $23.42_{-0.22}^{+0.20}$ & $23.48_{-0.32}^{+0.20}$ \\
&   ${\rm log}~(N_{\rm H22}/{\rm cm}^{-2})$ &  $23.26_{-0.18}^{+0.20}$ &  $22.98_{-0.22}^{+0.26}$ & $23.27_{-0.20}^{+0.28}$& &  $23.53_{-0.17}^{+0.16}$ & $<23.5$ &  $23.40_{-0.36}^{+0.20}$\\

&  $\lip_2$  & $3.28_{-0.18}^{+0.28}$ & $3.33_{-0.16}^{+0.12}$ & $3.48_{-0.26}^{+0.20}$ & & $3.42_{-0.17}^{+0.16}$ &   $3.46_{-0.13}^{+0.14}$  &   $3.52_{-0.30}^{+0.25}$  \\
&  $v_{\rm min}^{(1)}/c$  & $0.12_{-0.02}^{+0.02}$ &  $0.20_{-0.03}^{+0.03}$ & $0.12_{-0.01}^{+0.01}$ & & $0.08_{-0.03}^{+0.02}$ & $0.25_{-0.02}^{+0.02}$ & $0.11_{-0.03}^{+0.04}$  \\
&  $v_{\rm max}^{(1)}/c$   &  $0.28_{-0.03}^{+0.03}$ & $0.29_{-0.04}^{+0.04}$ & $0.15_{-0.02}^{+0.02}$ & &  $0.30_{-0.05}^{+0.05}$  &  $0.35_{-0.02}^{+0.02}$ & $0.26_{-0.08}^{+0.08}$  \\
&  $v_{\rm min}^{(2)}/c$  & $0.39_{-0.06}^{+0.06}$  &  $0.41_{-0.05}^{+0.05}$ &  $0.23_{-0.06}^{+0.06}$ & & $0.36_{-0.04}^{+0.04}$ & $0.48_{-0.05}^{+0.05}$ & $0.39_{-0.08}^{+0.08}$\\
&  $v_{\rm max}^{(2)}/c$   &  $0.49_{-0.02}^{+0.03}$ &  $0.50_{-0.02}^{+0.02}$ & $0.48_{-0.05}^{+0.05}$  & & $0.54_{-0.03}^{+0.03}$ & $0.58_{-0.06}^{+0.04}$ & $0.58_{-0.07}^{+0.07}$\\
& \lLa  & 46.95$\pm$0.08   & 46.92$\pm$0.07 & 46.91$\pm$0.08 & & 47.08$\pm$0.09 & 46.87$\pm$0.07 & 46.98$\pm$0.12 \\
&   $\chi^2/\nu$  & 55.1/65  & 65.5/58 & 67.4/63 & & 52.8/63 & 83.8/58 & 41.3/63\\
&   $P(\chi^2/\nu)$  & 0.81  & 0.23 & 0.33 & & 0.82  & 0.15 & 0.98\\ \\
\enddata

\tablenotetext{a}{MODEL1 is a power-law with Galactic, neutral,  and a one component outflow absorption (\xspec\ model {\sc wabs*zwabs*mtable\{windfull.tab\}*pow}); 
MODEL2 is a power-law with Galactic, neutral, and a two component outflow absorption (\xspec\ model {\sc wabs*zwabs*mtable\{windvslow.tab\}*mtable\{windvfast.tab\}*pow}).}

\tablenotetext{b}{The spectra fitted are the added (with {\sc heasoft} ftools routine
{\sc addspec}) spectra of the FI chips (XIS0, XIS2 and XIS3). For OBS 2
the XIS2 CCD was not operational, and only the XIS0 and XIS3
spectra were added for this observation. The BI spectra are taken
with the XIS1 chip.}

\tablenotetext{b}{\hbox{Logarithm} of the intrinsic (unabsorbed) X-ray luminosity in the 2-10~keV \RF. These luminosities have been corrected for absorption (both Galactic and intrinsic).}
\end{deluxetable*} 



\begin{deluxetable*}{cccccccccccc} \tabletypesize{\scriptsize}
\tablecolumns{12} \tablewidth{0pt} \tablecaption{
Results from
spectral fits of the \chandra\ and \xmm\ observations of \apm.
\label{tab:mocx}}

\tablehead{
\multicolumn{1}{c}{Model$^a$} & \multicolumn{1}{c}{Parameter} & \multicolumn{1}{c}{Values Epoch 1} & \multicolumn{1}{c}{Values Epoch 2} & \multicolumn{1}{c}{Values Epoch 3}  & \multicolumn{1}{c}{Values Epoch 4} & \multicolumn{1}{c}{Values Epoch 5} \\
}

\startdata

MODEL1...   &$\Gamma$  &  $1.75_{-0.03}^{+0.04}$   & $2.00_{-0.05}^{+0.06}$  &  $2.13_{-0.05}^{+0.09}$ &  $2.38_{-0.08}^{+0.09}$  &  $1.96_{-0.03}^{+0.03}$\\
&  ${\rm log}~(N_{\rm H1}/{\rm cm}^{-2})$ & $22.67_{-0.07}^{+0.04}$ & $22.72_{-0.07}^{+0.04}$   &  $22.71_{-0.02}^{+0.02}$ &  $22.68_{-0.02}^{+0.02}$ &  $23.02_{-0.03}^{+0.03}$\\

&   ${\rm log}~(N_{\rm H2}/{\rm cm}^{-2})$ &  $23.62_{-0.06}^{+0.05}$ &  $23.67_{-0.06}^{+0.06}$ & $23.76_{-0.05}^{+0.08}$ & $23.85_{-0.05}^{+0.06}$ & $23.13_{-0.11}^{+0.09}$ \\

&  $\lip_2$  & $3.56_{-0.12}^{+0.08}$ & $3.35_{-0.10}^{+0.09}$  & $3.47_{-0.10}^{+0.10}$  & $3.36_{-0.05}^{+0.05}$ & $3.28_{-0.10}^{+0.12}$ \\

&  $v_{\rm min}/c$  & $0.15_{-0.01}^{+0.01}$ & $0.13_{-0.01}^{+0.01}$  & $0.08_{-0.02}^{+0.02}$  &  $0.05_{-0.01}^{+0.01}$  &  $0.13_{-0.03}^{+0.03}$\\
&  $v_{\rm max}/c$   &  $0.43_{-0.04}^{+0.03}$ &  $0.64_{-0.06}^{+0.07}$ & $0.69_{-0.03}^{+0.03}$ & $0.73_{-0.02}^{+0.03}$ & $0.31_{-0.02}^{+0.02}$ \\
&  $p$   &  $3.8_{-1.0}^{+1.0}$  &$2.2_{-0.5}^{+0.8}$  & $1.6_{-0.2}^{+0.2}$ & $1.5_{-0.1}^{+0.2}$ & $1.2_{-0.4}^{+0.6}$ \\
& \lLa  & 46.83$\pm$0.06 & 47.01$\pm$0.07 & 47.15$\pm$0.08 & 47.48$\pm$0.12 & 46.91$\pm$0.04    \\
&   $\chi^2/\nu$  & 132.1/104 & 110.5/115 &109.4/105  & 152.4/143  & 77.2/70 \ \\
&   $P(\chi^2/\nu)$  & 0.03 &  0.60 & 0.36 & 0.28 & 0.26\\
\\

MODEL2...   &$\Gamma$  &  $1.77_{-0.02}^{+0.02}$  & $2.04_{-0.05}^{+0.05}$ & $2.17_{-0.07}^{+0.06}$ & $2.31_{-0.05}^{+0.08}$ &  $1.99_{-0.04}^{+0.05}$ \\
&    ${\rm log}~(N_{\rm H1}/{\rm cm}^{-2})$ &  $22.61_{-0.05}^{+0.05}$ &  $22.63_{-0.05}^{+0.05}$ &  $22.67_{-0.05}^{+0.04}$ &  $22.63_{-0.03}^{+0.03}$ & $23.01_{-0.04}^{+0.04}$  \\
&    ${\rm log}~(N_{\rm H21}/{\rm cm}^{-2})$ & $23.27_{-0.08}^{+0.11}$ & $23.17_{-0.13}^{+0.12}$ & $23.60_{-0.08}^{+0.07}$ & $23.34_{-0.10}^{+0.06}$ & $23.22_{-0.10}^{+0.08}$ \\
&    ${\rm log}~(N_{\rm H22}/{\rm cm}^{-2})$ & $23.08_{-0.12}^{+0.12}$ & $23.47_{-0.13}^{+0.10}$ & $23.41_{-0.12}^{+0.12}$ &  $23.51_{-0.07}^{+0.07}$ & $23.12_{-0.22}^{+0.13}$\\
&  $\lip_2$  & $3.32_{-0.19}^{+0.18}$ & $3.22_{-0.06}^{+0.06}$ & $3.37_{-0.08}^{+0.08}$ & $3.28_{-0.04}^{+0.04}$   & $3.25_{-0.11}^{+0.15}$   \\
&  $v_{\rm min}^{(1)}/c$  & $0.15_{-0.01}^{+0.01}$ & $0.13_{-0.02}^{+0.02}$ & $0.04_{-0.02}^{+0.02}$ & $0.06_{-0.03}^{+0.02}$ & $0.11_{-0.01}^{+0.01}$\\
&  $v_{\rm max}^{(1)}/c$   &  $0.17_{-0.01}^{+0.01}$ & $0.31_{-0.04}^{+0.04}$ &  $0.40_{-0.05}^{+0.05}$ & $0.21_{-0.05}^{+0.05}$ &  $0.31_{-0.02}^{+0.02}$\\
&  $v_{\rm min}^{(2)}/c$  & $0.32_{-0.02}^{+0.02}$  & $< 0.12$ & $0.43_{-0.10}^{+0.10}$ &  $0.24_{-0.06}^{+0.06}$ & $0.58_{-0.04}^{+0.04}$\\
&  $v_{\rm max}^{(2)}/c$   &  $0.40_{-0.01}^{+0.01}$ &  $0.63_{-0.03}^{+0.04}$ &  $0.69_{-0.03}^{+0.03}$ & $0.73_{-0.02}^{+0.02}$  &  $0.66_{-0.03}^{+0.02}$\\
& \lLa  & 46.79$\pm$0.05 & 47.03$\pm$0.06 &  47.18$\pm$0.08 &  47.40$\pm$0.07 &  46.98$\pm$0.06    \\
&   $\chi^2/\nu$  & 116.8/102 & 107.5/113& 106.5/103  & 152.5/141  &  66.2/68\\
&   $P(\chi^2/\nu)$  & 0.15 & 0.63 & 0.39 & 0.24 & 0.54 \\ \\

\enddata

\tablenotetext{a}{MODEL1 is a power-law with Galactic, neutral,  and a one component outflow absorption (\xspec\ model {\sc wabs*zwabs*mtable\{windfull.tab\}*pow}); 
MODEL2 is a power-law with Galactic, neutral, and a two component outflow absorption (\xspec\ model {\sc wabs*zwabs*mtable\{windvslow.tab\}*mtable\{windvfast.tab\}*pow}).}

\end{deluxetable*}

 
In several observations, the X-ray BALs produced by highly ionized iron lines (above 7~keV \RF) appear to consist of two broad absorption components.
The component covering low energies ($\lesssim$ 9 keV rest-frame) is referred to as slow 
and the one covering larger energies ($\gtrsim$ 10 keV rest-frame) is referred to as fast. Typical models used to fit
the  fast component contain a number of gaussian lines \citep{Cha02, Cha07a, Cha09a, Sae09}.
We created the table model {\sc windvslow.tab} to fit the slow component
and the table model {\sc windvfast.tab} to fit the fast component.
Since table models {\sc windvslow.tab} and {\sc windvfast.tab} are used to fit a relative small 
portion of the spectra we fix the velocity profile parameter to $p=1.0$.  
Each component of the outflow has a velocity gradient ${\Delta v}$ with a minimum ($v_{\rm min}$)
and maximum velocity ($v_{\rm max}$) that satisfy $v_{\rm max}=v_{\rm min}+\Delta v$. By fitting these table models to the spectra we constrain the minimum and maximum projected velocities of each component.
Table model windvslow.tab is generated with $\Delta v$ having values in the range 0.02-0.42$c$  (step intervals of 0.04$c$) and the parameters ($v_{\rm min}$), $\lnh$ and $\lip$ cover the same range and values as those used for producing table model {\sc windfull.tab}.
Table model windvfast.tab is generated with $\Delta v$ having values in the range 0.04-0.80$c$  (step intervals of 0.04$c$ for $\Delta v < 0.24c$ and step intervals of 0.08$c$ for $\Delta v \geq 0.24c$) and the parameters $v_{\rm max}$, $\lnh$ and $\lip$ cover the same range and values as those used for producing table model {\sc windfull.tab}.

\subsubsection{Results from fits to X-ray spectra of \apm} \label{S:xfit}

We fit the X-ray spectra of \apm\ using two different models.
These two models assume a source spectrum consisting of a power-law attenuated by Galactic absorption (\xspec\ model {\sc wabs}).
We assumed a Galactic column density of $4.1 \times 10^{20} \cmsq$ \citep{Kal05}.
The first model (MODEL1) contains an intrinsic neutral absorber (\xspec\ model {\sc zwabs}) 
to describe the absorption in the low energy range of 1$-$4~keV \RF\ and 
an outflowing ionized absorber (\xspec\ table model {\sc windfull.tab}) to account for the X-ray BALs. In \xspec\ notation MODEL1 is written as:\\ {\sc wabs*zwabs*mtable\{windfull.tab\}*pow}.
The second model (MODEL2) contains an intrinsic neutral absorber (\xspec\ model {\sc zwabs})  to describe the absorption in the low energy range of 1$-$4~keV \RF\ and 
a two component outflowing ionized absorber (\xspec\ table models {\sc windvslow.tab} and {\sc windvfast.tab}) 
to account for the X-ray BALs. In \xspec\ notation MODEL1 is written as:\\
 {\sc wabs*zwabs*mtable\{windvslow.tab\}*\\mtable\{windvfast.tab\}*pow}.

We note that MODEL1 and MODEL2 use a neutral absorber to describe the intrinsic attenuation in the low energy range of 1$-$4~keV \RF. In this work we also tried fits with an ionized absorber to describe the attenuation at 1$-$4~keV; these fits did not result in a significant  improvement over fits that used a neutral absorber.  Additionally, as discussed in \cite{Sae09, Cha09a}, the use of a neutral absorber to describe the absorption in the low energy range of 1$-$4~keV \RF\  did not improve with the use of more complex absorbers 
even for spectral fits performed to deep \XR\ observations of \apm.  We also note that given the high redshift of \apm,  our fits to the spectra of this source cannot adequately constrain the low-energy intrinsic  absorption at 1$-$4~keV \RF. \\
For MODEL2 the current data cannot adequately constrain the ionization parameters
of both the slow and fast components of the outflow. We therefore set the ionization parameters of the slow and fast component to be equal in MODEL2.\footnote{Notice that in \cite{Sae09} we tried a similar model to MODEL2 (model XSTAR4, Table~5) and we found that the ionization state of the slow component could not be distinguished for the ionization state of the fast component in the \suzaku\ observations. We have a similar case when we try MODEL2 on our eight X-ray observations without setting the ionization parameter of the slow and fast component as equal.}


 \begin{deluxetable*}{ccccc} \tabletypesize{\small}
\tablecolumns{11} \tablewidth{0pt} \tablecaption{ Estimates of the
improvement of fits to the spectra of APM~08279+5255
using MODEL2 over MODEL1. \label{tab:ftes}}

\tablehead{
 \colhead{OBSID} & \colhead{Instrument} &
\colhead{$F$-statistic$^a$} & \colhead{null probability} & \colhead{significance}}

\startdata

OBS1  & FI & 2.59 & 4.8 $\times$ 10$^{-2}$ & 95.2\%\\
OBS1  & BI & 3.05  & 1.5 $\times$ 10$^{-1}$ & 85.5\% \\
OBS2  & FI & 5.32 & 2.4 $\times$ 10$^{-1}$ & 76.0\%\\
OBS2  & BI & 7.15  & 1.4 $\times$ 10$^{-1}$  & 85.6\%\\
OBS3 & FI  & 1.31  & 2.8 $\times$ 10$^{-1}$ & 72.3\%\\
OBS3 & BI & 1.91 & 1.6 $\times$ 10$^{-1}$ & 84.3\%\\
Epoch~1 & ACIS S3 & 6.68 & 1.8 $\times$ 10$^{-3}$ & 99.8\% \\
Epoch~2 & EPIC pn  & 1.56 & 2.2 $\times$ 10$^{-1}$ & 78.4 \%\\
Epoch~3 & EPIC pn & 1.40 & 2.5 $\times$ 10$^{-1}$ & 74.9 \%\\
Epoch~4 & EPIC pn & ... & ... & ...\\
Epoch~5 & ACIS S3 & 5.66 & 5.32 $\times$ 10$^{-3}$ & 99.5\% \\

\enddata

\tablenotetext{a}{The value on the left of the slash is the $F$-statistic and is given by
$F=\frac{\chi_{\nu_1}^2-\chi_{\nu_1}^2}{\Delta
\nu}/\frac{\chi_{\nu_2}^2}{\nu_2}$. The value on the right of the slash
represents the probability of exceeding the $F$-statistic based on the $F$-test.}

\end{deluxetable*}


The results of the fits of these models to the \suzaku\ spectra of \apm\ are presented in 
Table~\ref{tab:mosu} and to the \chandra\ and \xmm\ spectra in Table~\ref{tab:mocx}.  
We note that in MODEL1 of Tables~\ref{tab:mosu} and \ref{tab:mocx}  $\lnh_1$ and $\lnh_2$ are the best fitted column densities of the intrinsic neutral absorber and the outflowing ionized absorber, respectively. In MODEL2 of these same tables $\lnh_1$, $\lnh_{21}$, $\lnh_{22}$ are the best fitted column densities of the intrinsic neutral absorber,  the slow and fast outflowing ionized absorber, respectively. In MODEL2 of these tables the velocity superscripts (1) and (2) represent the best-fitted velocity parameters of the slow and fast component, respectively.

In general the models that fit the high-energy outflowing absorber (see Tables~\ref{tab:mosu} and \ref{tab:mocx})  tend to have ionization parameters in the range $3.2 \lesssim \lip \lesssim 3.7$. With the exception of the \chandra\ observations, there is no improvement in the fits when using MODEL2 over MODEL1. In Table~\ref{tab:ftes} we show the statistical 
improvements based on the $F$-test of fits using MODEL2 over those using MODEL1. For the \chandra\ observations the improvements are $\gtrsim 99.5\%$ when we use MODEL2 over MODEL1.  There are also marginal improvements using MODEL2 over MODEL1 for the  \suzaku\ observations especially in OBS1 (see Table~\ref{tab:ftes}). In general we will use MODEL2 to estimate the outflow parameters because this model provides better constrains of the maximum velocity of the outflow. However, in the case of the \xmm\ observations we use MODEL1 to estimate the outflow parameters because these observations show the widest X-ray BALs and provide better constraints of the velocity profile parameter $p$.   We mainly used MODEL 1 to fit the \xmm\ spectra of \apm\ because only one broad absorption trough is clearly observed at high energies.
When we used Model 2 to fit the \xmm\ spectra of \apm\ the
two outflow components were found to overlap in velocity space. 
This overlap occurs when values of $v_{\rm max}^{(1)}$ are less or similar to $v_{\rm min}^{(2)}$, as presented in Table~\ref{tab:mocx}.

Based on the spectral fits to the \suzaku\ observations (Table~\ref{tab:mosu}), we find possible variability   ($\gtrsim 2\sigma$)  in the minimum velocity of the outflow between OBS2 and OBS3 ($v_{\rm min}$ in MODEL1 and $v_{\rm min}^{(1)}$ in MODEL 2) for both the BI and FI spectra.  The  evidence of this variability in the \suzaku\ spectra of \apm\  was already  analyzed in \cite{Sae09} and in this work we conclude that its significance is at the 99.9\% and 98\% significance levels in the FI and BI spectra, respectively.  
This variability is also confirmed  to exist between the column density and velocity of the slow component when we compare epochs OBS2 and OBS3.

\begin{figure}
\includegraphics[width=6.2cm]{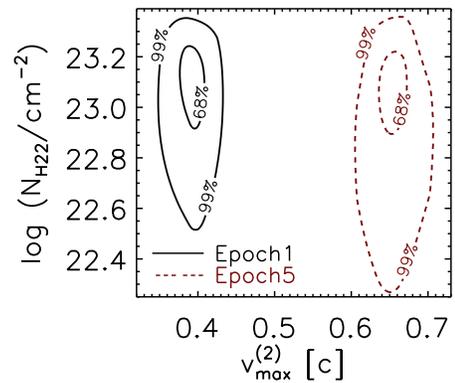}
        \centering
    \caption[Contour plots of the column density (${\rm log}~(N_{\rm H22}/{\rm cm}^{-2})$) versus maximum velocity   ($v_{\rm max}^{(2)}$) of the fast component of the ionized outflow absorber.]{Contour plots of the column density (${\rm log}~(N_{\rm H22}/{\rm cm}^{-2})$)  versus maximum velocity   ($v_{\rm max}^{(2)}$) of the fast component of the ionized outflow absorber. The contour plots have been calculated  for the \chandra\ observations Epoch~1 (continuous line) and Epoch~5 (dashed line) at the 68\% and 99\% level of significance using MODEL2 of Table~\ref{tab:mocx}. }
      \label{fig:coch}
\end{figure}

We confirm the dramatic change in the maximum velocity of the fast component of the \chandra\ observations as shown in \cite{Cha09a}. We find a $\gtrsim 3  \sigma$ change of $v_{\rm max}^{(2)}$ between \hbox{Epoch 1} and Epoch 5 using either MODEL1 or MODEL2 (see Table~\ref{tab:mocx}). 
We also confirm this change in velocity when we plot the confidence contours of the column density ($\lnh_{22}$) versus the maximum velocity of the fast outflow component ($v_{\rm max}^{(2)}$) using MODEL2. In Figure~\ref{fig:coch} we present confidence contours of $\lnh_{22}$ versus $v_{\rm max}^{(2)}$ for Epoch 1 and Epoch 5. The significance of the change of  $\lnh_{22}$ versus $v_{\rm max}^{(2)}$ between Epochs 1 and 5  is $\gtrsim 99.9 \%$ with a null probability of $\lesssim 10^{-4}$.

\begin{figure}
\includegraphics[width=8.6cm]{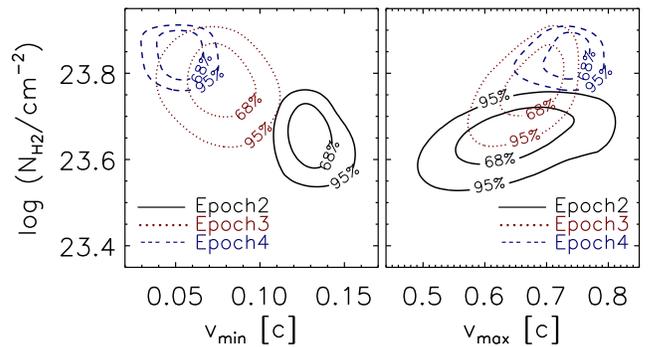}
        \centering
    \caption[Confidence contours of the total column density  (${\rm log}~(N_{\rm H2}/{\rm cm}^{-2})$)  versus minimum velocity ($v_{\rm min}$; left panel) and maximum velocity ($v_{\rm max}$; right panel)   of  the high-energy absorber.]{Confidence contours of the total column density  (${\rm log}~(N_{\rm H2}/{\rm cm}^{-2})$) versus minimum velocity ($v_{\rm min}$; left panel) and maximum velocity ($v_{\rm max}$; right panel)   of  the high-energy absorber. The contour plots have been calculated for the \xmm\ observations Epoch~2 (solid line), Epoch~3 (dotted line) and Epoch~4 (dashed line) at the 68\% and 95\% level of significance using MODEL1 of Table~\ref{tab:mocx}. }
      \label{fig:coxm}
\end{figure}

Using either MODEL1 or MODEL2 we find a possible change in the maximum velocity of the outflow  between the \xmm\ observations of Epoch 2 and Epoch 4, however, the significance of this change is only at the 1$-$$\sigma$ level (see Table~\ref{tab:mocx}). We do find however significant changes in the minimum velocity and total column density of the outflow ($\lnh_2$; Table~\ref{tab:mocx})  between Epochs 2 and 4.  
In Table~\ref{tab:mocx} we show that this variability is significant at the 
$\gtrsim$~2$-$$\sigma$  level for fits using MODEL1. In Figure~\ref{fig:coxm} we plot confidence contours of
$\lnh_2$ versus $v_{\rm min}$ (left panel) and $ \lnh_2$   versus $v_{\rm max}$ (right panel) for each \xmm\ observation for fits using MODEL1.   
We find that the significance of the variability  between Epoch~2 and Epoch~4  of $\lnh_2$  versus $v_{\rm min}$  is significant at the $\gtrsim 99.9\%$ level and of $\lnh_2$  versus $v_{\rm max}$  is significant at the $\gtrsim 99\%$ level.
We also find variability  in the minimum velocity of the outflow when we compare Epoch~2 and Epoch~3. 
This can be seen in Table~\ref{tab:mocx} and the left panel of Figure~\ref{fig:coxm}.  
From the confidence contours  (Figure~\ref{fig:coxm}, left panel), we estimate the significance 
of the variability on  $\lnh_2$  versus $v_{\rm min}$  between 
Epoch 2 and Epoch 3 to be $\gtrsim 99\%$.  It is worth mentioning that the fits using MODEL2 show an increase in the maximum velocity $\sim$~2$-$$\sigma$ of the slow outflow component ($v_{\rm max}^{(1)}$) between Epoch~3 and Epoch~4.  This change is consistent with the change found in the first component energy  of the two gaussian absorption line model of \cite{Cha09a}.  
Finally we note that the spectral fits to the \xmm\  observations show values of $p$ that are greater than one. Although we find indications of variability of $p$, these changes are not significant given the poor constraints of this parameter.

\begin{figure}
\includegraphics[width=8.6cm]{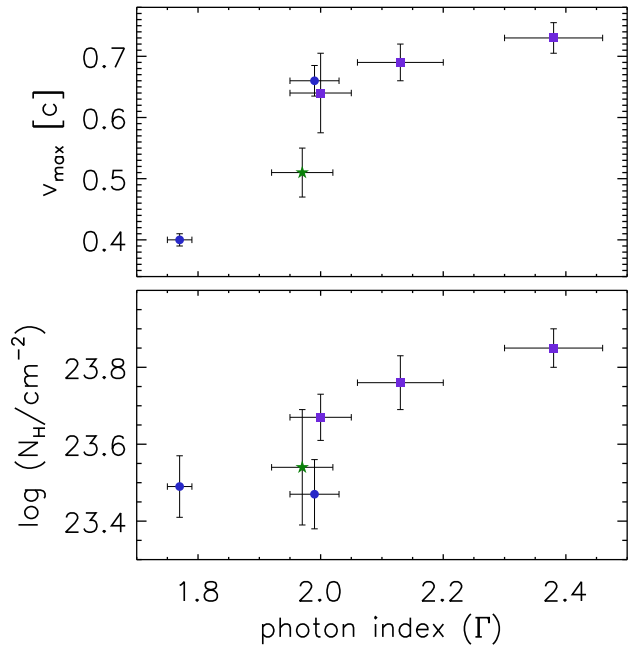}
        \centering
    \caption[Maximum velocity  ($v_{\rm max}$; upper panel) and total column density of the outflow (\lnhcm; lower panel) versus photon index ($\Gamma$) for our eight deep observations.]{Maximum velocity  ($v_{\rm max}$; upper panel) and total column density of the outflow (\lnhcm; lower panel) versus photon index ($\Gamma$) for our eight deep observations.
 $v_{\rm max}$, \lnhcm\ and $\Gamma$ were derived from fits to the spectra of 
APM 08279+5255 with a model that included an absorbed power-law and two component outflow (MODEL2 of Tables~\ref{tab:mosu} and \ref{tab:mocx}) for the \suzaku\ and \chandra\ observations and a single component outflow for the \xmm\ observations (MODEL1 of Table~\ref{tab:mocx}). Errors shown are at the 68\% confidence level.  In each panel the \suzaku\ data points are displayed as a single point that represents the weighted (by total counts) mean of the  three \suzaku\ observations.  Data  shown with circles, a star, and squares are obtained from \chandra, \suzaku\ and \xmm\ observations respectively.}
      \label{fig:vnGa}
\end{figure}

\subsubsection{Mass outflow rate and kinetic energy of outflow of \apm}

We use results from fits of our quasar outflow model to the spectra of APM 08279+5255 to constrain the 
mass outflow rate and kinetic energy injected by this outflow into the 
surrounding medium. In \cite{Sae09} and \cite{Cha09a} we also constrained these properties of the outflow,
however, these estimates were based on less realistic models (see \S \ref{S:Xobs}).


\begin{deluxetable*}{cccccccccccc} \tabletypesize{\scriptsize}

\tablecolumns{12} \tablewidth{0pt} \tablecaption{
Mass-outflow rates
and efficiencies of outflows in \apm\ ${}^{a}$.
\label{tab:Moek}}

\tablehead{
\multicolumn{1}{c}{OBS} & \multicolumn{1}{c}{Instr.}  & \multicolumn{1}{c}{$\dot{M}$ (abs1)} & \multicolumn{1}{c}{$\epsilon_K$ (abs1)} &\multicolumn{1}{c}{$\dot{M}$ (abs2)}  & \multicolumn{1}{c}{$\epsilon_K$ (abs2)} & \multicolumn{1}{c}{$\dot{M}$ (tot)$^b$}  & \colhead{$\epsilon_K$ (tot)$^b$}   \\
& &   \multicolumn{1}{c}{[$10^3 M_\odot \mu_L^{-1}{\rm yr}^{-1}$]} & & \multicolumn{1}{c}{[$10^3 M_\odot \mu_L^{-1}{\rm yr}^{-1}$]}  & &\multicolumn{1}{c}{[$10^ 3 M_\odot \mu_L^{-1}{\rm yr}^{-1}$]} \\  
}

\startdata

OBS1 & XIS FI  &  $1.0^{+0.9}_{-0.6}$ & $0.05^{+0.05}_{-0.03}$ &  $2.2^{+2.2}_{-1.5}$ & $0.6^{+0.7}_{-0.4}$ & $3.2^{+3.2}_{-2.1}$/$3.5^{+3.4}_{-2.3}$ &  $0.6^{+0.7}_{-0.4}$/$0.9^{+1.0}_{-0.6}$ \\

OBS2 & XIS FI  & $0.9^{+0.8}_{-0.6}$ & $0.06^{+0.07}_{-0.04}$ &  $1.0^{+1.0}_{-0.7}$ & $0.3^{+0.3}_{-0.2}$ & $1.9^{+1.9}_{-1.2}$/$2.0^{+2.0}_{-1.4}$ &  $0.3^{+0.4}_{-0.2}$/$0.5^{+0.5}_{-0.3}$ \\

OBS3 & XIS FI  & $4.6^{+4.4}_{-3.0}$ & $0.01^{+0.01}_{-0.01}$ &  $1.8^{+1.8}_{-1.2}$ & $0.3^{+0.4}_{-0.2}$ & $2.3^{+2.2}_{-1.5}$/$2.5^{+2.5}_{-1.7}$ &  $0.4^{+0.4}_{-0.2}$/$0.5^{+0.6}_{-0.3}$  \\

OBS1 & XIS BI  & $1.7^{+1.7}_{-1.1}$ & $0.11^{+0.13}_{-0.08}$ &  $4.2^{+4.1}_{-2.7}$ & $1.1^{+1.2}_{-0.7}$ & $5.9^{+5.8}_{-3.9}$/$6.5^{+6.4}_{-4.2}$ &  $1.2^{+1.3}_{-0.8}$/$1.7^{+1.8}_{-1.1}$ \\

OBS2 & XIS BI  & $2.2^{+2.2}_{-1.5}$ & $0.24^{+0.26}_{-0.16}$ &  $2.7^{+2.6}_{-1.9}$ & $1.1^{+1.3}_{-0.7}$ & $4.8^{+4.9}_{-3.3}$/$5.3^{+5.3}_{-3.7}$ &  $1.3^{+1.5}_{-0.9}$/$1.8^{+2.2}_{-1.3}$ \\

OBS3 & XIS BI  & $1.6^{+1.6}_{-1.1}$ & $0.09^{+0.14}_{-0.07}$ &  $3.4^{+3.3}_{-2.3}$ & $1.2^{+1.5}_{-0.9}$ & $5.0^{+5.0}_{-3.4}$/$5.4^{+5.4}_{-3.7}$ &  $1.3^{+1.6}_{-0.9}$/$1.9^{+2.4}_{-1.4}$  \\

Epoch 1 & ACIS S3 & $0.8^{+0.8}_{-0.5}$ & $0.02^{+0.02}_{-0.02}$ &  $1.2^{+1.1}_{-0.8}$ & $0.2^{+0.2}_{-0.1}$ & $2.1^{+1.9}_{-1.3}$/$2.2^{+2.1}_{-1.4}$ &  $0.2^{+0.2}_{-0.1}$/$0.3^{+0.3}_{-0.2}$  \\

Epoch2 & EPIC pn & ... & ... & ... & ... &  $3.9^{+3.6}_{-2.5}$/$4.3^{+4.0}_{-2.7}$ &  $0.9^{+1.0}_{-0.6}$/$1.4^{+1.6}_{-1.0}$ \\

Epoch3 & EPIC pn & ... & ... & ... & ... &  $5.2^{+4.9}_{-3.3}$/$5.8^{+5.4}_{-3.7}$ &  $1.6^{+1.5}_{-1.0}$/$2.5^{+2.5}_{-1.6}$ \\

Epoch4 & EPIC pn & ... & ... & ... & ... &  $6.5^{+5.8}_{-4.1}$/$7.4^{+6.6}_{-4.7}$ &  $2.3^{+2.1}_{-1.5}$/$4.0^{+3.7}_{-2.6}$ \\

Epoch5 & ACIS S3 & $0.9^{+0.8}_{-0.6}$ & $0.05^{+0.05}_{-0.03}$ &  $2.3^{+2.3}_{-1.5}$ & $1.4^{+1.4}_{-0.9}$ & $3.2^{+3.1}_{-2.1}$/$3.5^{+3.4}_{-2.3}$ &  $1.5^{+1.4}_{-1.0}$/$2.3^{+2.3}_{-1.5}$  \\

\enddata

\tablenotetext{a}{Estimated values of the outflow properties of \apm\ 
based on a model that includes an absorbed power-law and a two component outflow (MODEL3 of Tables~\ref{tab:mosu} and \ref{tab:mocx}) for the \suzaku\ and \chandra\ observations and a single component outflow for the \xmm\ observations (MODEL1 of Table~\ref{tab:mocx}). The values of $\dot{M}$ and $\epsilon_K$ are
obtained from equations (\ref{eq:Mdot}) and (\ref{eq:ek}) assuming $M_{\rm
BH}\sim 10^{12}\mu_L^{-1}M_\odot$ (see \S\ref{S:gapm}) and $L_{\rm
bol}=7\times10^{15}\mu_L^{-1}L_\odot$ \citep{Irw98,Rie09}.}

\tablenotetext{b}{We provide two estimates of  $\dot{M}$(tot) and $\epsilon_K$(tot). The estimate on the left 
 corresponds to the case where the angle between our line of sight and the outflow direction is zero
 while the one on the right corresponds to an angle of 10$^\circ$.}

\end{deluxetable*}


A trend between $v_{\rm max}$ and $\Gamma$ based on the X-ray observations of \apm\ (Figure~\ref{fig:vnGa}, upper panel) was 
recently found by \cite{Cha09a}. Such a trend is consistent with models that predict that  quasar winds are driven by radiation pressure.
Our fits with a more realistic quasar outflow model confirm the trend between $v_{\rm max}$ and $\Gamma$.
We also find a possible trend between the total column density of the outflowing ionized absorber (\lnh) and  $\Gamma$. 
As the lower panel of Figure~\ref{fig:vnGa} shows, we find that \lnh\ increases with $\Gamma$.  The increase of the maximum velocity with 
$\Gamma$ could be related to the fact that softer spectra have a stronger radiative effect on the wind \citep[see \S \ref{S:SEDs} and \S 3.3 of][]{Cha09a}.
On the other hand, a possible increase in column density with $\Gamma$ could be indicating that \apm\ becomes a more effective radiative driving source as the spectrum of the ionizing source becomes softer. 
Additional X-ray observations of \apm\  will show if these trends are significant. 
 
In order to calculate the mass outflow rate ($\dot{M}$) and the efficiency of the wind ($\epsilon_K$) we have modified the formulas used in \cite{Sae09} and \cite{Cha09a} to include the modeled velocity gradient of the wind and include special relativistic corrections.
The mass outflow rate is given by ($\dot{M}=4 \pi f_c m_p  \frac{R^2}{\Delta R}  \int_0^{\nh}v(\nhat)d \nhat$)

\begin{equation}  \label{eq:Mdot}
\dot{M}=4 \pi f_c   \frac{R^2}{\Delta R} \nh m_p \left(v_{\rm min}+\frac{v_{\rm max}-v_{\rm min}}{p+1} \right), 
\end{equation}  

and the wind efficiency, defined as the ratio of the rate of kinetic energy
injected into the interstellar medium and IGM by the outflow to
the quasarÕs bolometric luminosity,  is ($\epsilon_K=\frac{c^2}{L_{\rm bol}}\int (\gamma-1) d \dot{M}$)

\begin{equation} \label{eq:ek}
\epsilon_K=\frac{4 \pi c^2 f_c m_p}  {L_{\rm bol}} \frac{R^2}{\Delta R} \int _0^{\nh}(\gamma-1)v(\nhat)d\nhat.
\end{equation}

In equations~(\ref{eq:Mdot}) and (\ref{eq:ek})  $f_c$ is the global covering factor, \nh\ is the column density, $R$ is the radius, and $\Delta R$ is the thickness of the absorber.  We note that for an absorber with constant non-relativistic speed ($v/c \ll 1$) equations   (\ref{eq:Mdot}) and (\ref{eq:ek}) have the same form as the ones used in \cite{Sae09} and \cite{Cha09a} 
\citep[see e.g., equation 4  of][]{Sae09}.\footnote{For a constant velocity, i.e. a $p$-type profile (equation \ref{eq:vNH}) with $p=0$ and $v_{\rm min} =v$, then equation (\ref{eq:Mdot}) reduces to \hbox{$\dot{M}=4 \pi f_c (R^2/{\Delta R})N_{\rm H} m_p v $}. Also, since $\gamma \approx1+v^2/(2c^2)$ ($v/c \ll 1$) equation (\ref{eq:ek}) becomes $\epsilon_K = \dot{M} v^2/(2 L_{\rm bol})$.}
The velocity profile assumed in our models is given by equation~(\ref{eq:vNH}),  therefore equations (\ref{eq:Mdot}) and (\ref{eq:ek}) contain the dependence of 
the estimated outflow properties on the model parameters of the velocity profile. 
To obtain error bars for $\epsilon_K$ and $\dot{M}$, we performed a Monte Carlo simulation, assuming a uniform distribution of the parameters $f_c$, $R$, and $R/\Delta R$ around the expected values of these parameters, and a normal distribution for \lnh, $v_{\rm min}$ and $v_{\rm max}$  (described by the parameters in Tables~\ref{tab:mosu} and \ref{tab:mocx}). Specifically, we assume a global covering factor lying in the range $f_c = 0.1-0.3$, based on the observed fraction of BAL quasars  \citep[e.g.,][]{Hew03,Gib09} and a fraction $R/\Delta R$ ranging from 1 to 10 based on current theoretical models of quasar outflows \citep{Pro00, Pro04}. Based on our estimated maximum velocities ($v_{\rm max} \sim 0.7c$) \citep[see e.g.,][]{Cha09a}  and the fast variability of the outflow we expect that $R$ will be close to the Schwarzschild radius ($R_S$).  Therefore in the Monte Carlo simulation we allow $R$ to vary between $3R_S$ and $15R_S$.

\begin{figure}
\includegraphics[width=7.6cm]{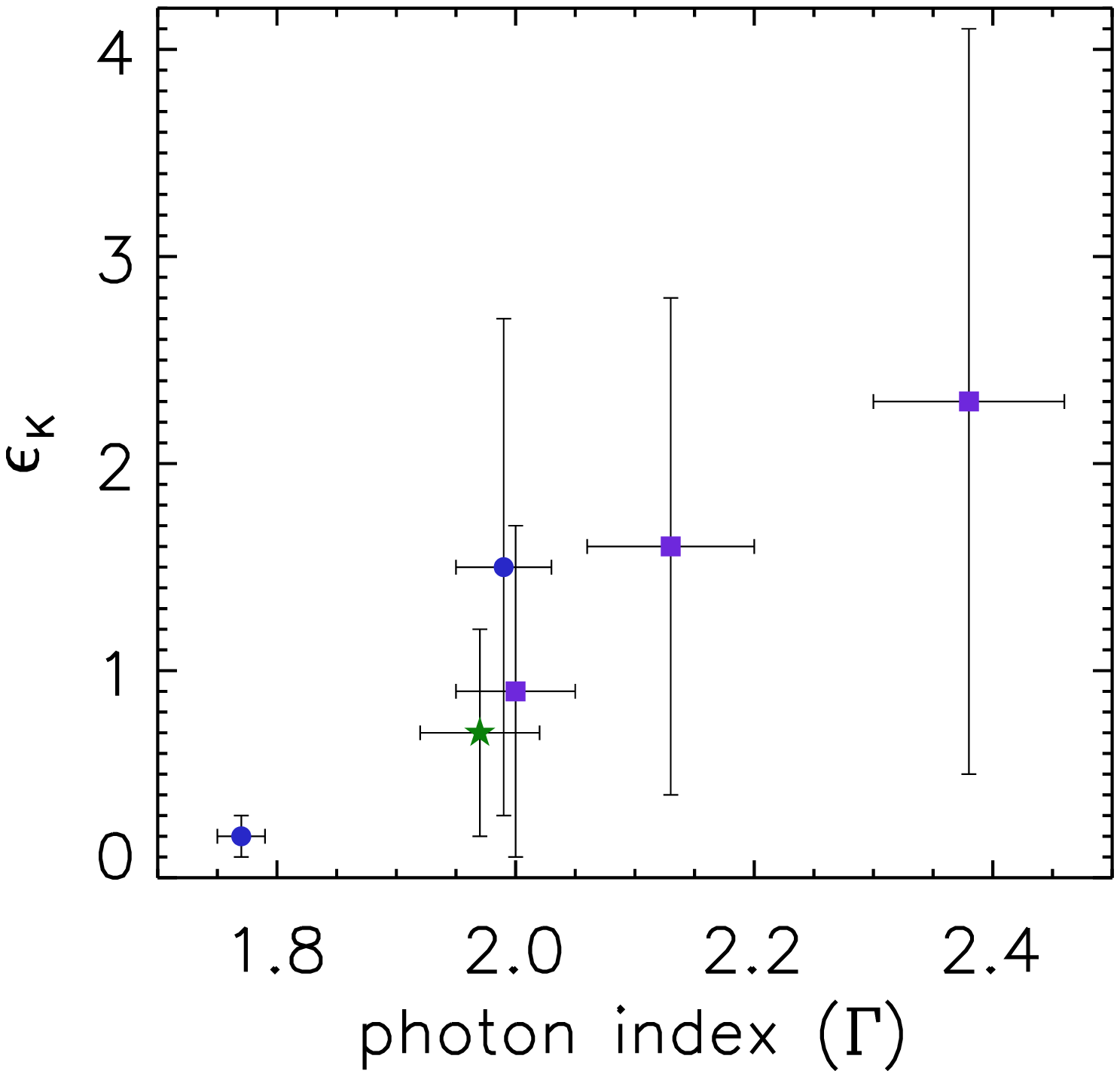}
        \centering
    \caption[Outflow efficiency ($\epsilon_K$) of the outflow  versus photon index ($\Gamma$) for  our eight  observations of \apm.]{Outflow efficiency ($\epsilon_K$) of the outflow  versus photon index ($\Gamma$) for  our eight  observations of \apm. $\epsilon_K$ and $\Gamma$  were derived from fits to the spectra of \apm\ with a model that included an absorbed power-law and two component outflow (MODEL2 of Tables~\ref{tab:mosu} and \ref{tab:mocx}) for the \suzaku\ and \chandra\ observations and a single component outflow for the \xmm\ observations (MODEL1 of Table~\ref{tab:mocx}). Errors shown are at the 68\%  confidence level. The \suzaku\ data points are displayed as a single point that represents the weighted (by total counts) mean of the  three \suzaku\ observations. Data  shown with circles, a star, and squares are obtained from \chandra, \suzaku\ and \xmm\ observations respectively.}
      \label{fig:ekGa}
\end{figure}

In Table~\ref{tab:Moek} we show the mass outflow rate ($\dot{M}$) and the efficiency ($\epsilon_K$) of the wind in each observation. 
The parameters used in equations~(\ref{eq:Mdot}) and (\ref{eq:ek})  to derive $\dot{M}$  and  $\epsilon_K$ were obtained from spectral fits of MODEL2 
to the \suzaku\ and \chandra\ observations, and through the use of MODEL1 for the \xmm\ observations (see Tables~\ref{tab:mosu} and \ref{tab:mocx}). For the \suzaku\ and \chandra\ observations we estimate $\dot{M}$ and $\epsilon_K$ for the slow and fast component of the outflow. 
The total mass outflow rate and efficiency is obtained by summing the contributions of the slow and fast outflow component (MODEL2, \S~\ref{S:xfit})  when multiple  components are required to fit the data. 
In the case of the \xmm\ observations no summing is required since 
a one component wind model (MODEL1, \S~\ref{S:xfit}) provides better fits (in a statistical sense) to the spectra of \apm\ (see \S~\ref{S:xfit}).  
In Table~\ref{tab:Moek}, assuming that the outflow is viewed along our line of sight, we show that during the observations of \apm\ the total mass outflow rate varied between $21^{+19}_{-13} - 65^{+58}_{-41} $~$M_{\odot}{\rm yr}^{-1}$ ($\mu_L=100$) and the total efficiency varied between $0.2^{+0.2}_{-0.1} - 2.3^{+2.1}_{-1.5} $.
However, the total values of $\dot{M}$ and $\epsilon_K$ are $\sim10\%$ and $\sim 50\%$ higher respectively if we assume an outflow forming an angle of $10^\circ$ with our line of sight (see Table~\ref{tab:Moek}).   In Table~\ref{tab:Moek}  we also show that an important fraction ($>10\%$) of the  bolometric energy of \apm\ is injected into the surrounding galaxy through quasar winds.

As Figure~\ref{fig:vnGa} shows the softer spectra probably result in a faster and more massive (higher total column density)  wind.
This tendency  can be seen more clearly in Figure~\ref{fig:ekGa} where we have plotted the total efficiency ($\epsilon_K$) as a function of the photon index ($\Gamma$). We note that a small contribution to the error bars ($\lesssim 20\%$) of $\epsilon_K$ arises from errors in the fitted parameters (i.e., $v_{\rm min}$, $v_{\rm max}$ and $\lnh$) while most of the error in $\epsilon_K$  is due to the uncertainty of $R$, $f_c$, $R/ \Delta R$.  

\begin{figure}
\includegraphics[width=8.6cm]{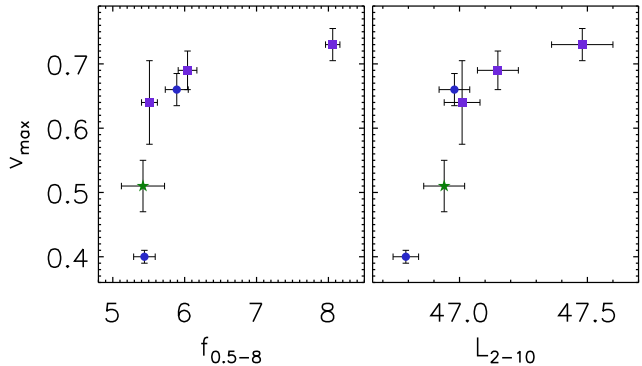}
        \centering
    \caption[]{Maximum velocity  ($v_{\rm max}$; upper panel) versus absorbed flux in the 0.5$-$8~keV band ($f_{0.5-8}$; left panel) and  intrinsic luminosity in the 2$-$10~keV \RF\  band ($L_{2-10}$; right panel) for our eight deep observations. $v_{\rm max}$ and $L_{2-10}$  were derived from fits to the spectra of \apm\ with a model that included an absorbed power-law and two component outflow (MODEL2 of Tables~\ref{tab:mosu} and \ref{tab:mocx}) for the \suzaku\ and \chandra\ observations and a single component outflow for the \xmm\ observations (MODEL1 of Table~\ref{tab:mocx}). The absorbed fluxes in the 0.5$-$8~keV band are from Table~\ref{tab:obid}. Errors shown are at the 68\%  confidence level. In each panel the \suzaku\ data points are displayed as a single point
that represents the weighted (by total counts) mean of the  three \suzaku\ observations. Data  shown with circles, a star, and squares are obtained from \chandra, \suzaku\ and \xmm\ observations respectively.}
        \label{fig:fvmx}
\end{figure}

The 0.5$-$8~keV flux of \apm\  varies by up to 50\% in the observations of our study
(see Table~\ref{tab:obid}). These flux changes appear to be correlated to changes of the maximum outflow 
velocity (see Figure~\ref{fig:fvmx}; left panel). 
The observed changes in flux may be caused by changes in the luminosity of the ionizing source. 
By fitting our outflow model to the X-ray spectra of \apm\ we obtain the intrinsic (unabsorbed) 
2$-$10~keV rest-frame X-ray luminosity of the ionization source ($L_{2-10}$) (see Tables~\ref{tab:mosu} and \ref{tab:mocx}). 
We find that the maximum velocity of the outflow is possibly correlated to
the intrinsic luminosity of the ionizing source (see Figure~\ref{fig:fvmx}; right panel). 
From the radiative outflow wind equation we obtain that the terminal velocity of the wind is
$v_{\infty} \propto \sqrt{\Gamma_f L}$, where  $L$ is the luminosity producing the radiative driving and  $\Gamma_f$ is the force multiplier 
(see \S4 for an analysis describing the dependence of $\Gamma_f$ with the photon index).
This equation predicts that an increase in the luminosity of the ionizing source should result in an increase of the 
terminal velocity of the outflow as observed.
The \suzaku\ observations of \apm\ do not show any significant changes of $\Gamma$, $v_{\rm max}$, the total column density, and $L_{2-10}$ (see Table~\ref{tab:mosu}).
In Figures~\ref{fig:vnGa}, \ref{fig:ekGa} and \ref{fig:fvmx} we therefore chose to represent the \suzaku\ observations with one data point.
The \suzaku\ data point in each figure is the weighted mean of the three \suzaku\ observations.

As a final comment, the ionization parameter obtained from the spectral fits is based on the assumption that the incident spectrum is a 
pure power-law with a fixed slope ($\alpha = -1$). In reality the spectra are likely more complex and variable than assumed in our model. 
Our simple approach provides reliable constrains on the column density and velocity of the outflow. However, since the ionization parameter depends on the SED from 1$-$1000 Ry (see Appendix \ref{ap:ionp}) we do not expect 
that fits of our quasar outflow model to the spectra of APM~08279 +5255 to constrain the ionization parameter since 
our model assumes a fixed SED. What our model does constrain is the temperature of the absorber which 
is a reliable indicator of the degree of ionization of the outflowing gas.  \footnote{ For example, for a fixed column density and velocity profile we have checked that approximately the same absorption features are generated if we use an incident \PL\ SED with $\alpha=-1.0$ and $\lip \approx 3.4$, an incident \PL\ with $\alpha=-0.6$ and $\lip \approx 2.7$, an incident \PL\ with $\alpha=-1.2$ and $\lip  \approx 3.8$, or an incident  \MF\ SED with $\lip \approx 4.7$. We note that the average temperatures of the gases producing these absorption profiles are approximately independent of the incident SEDs and lie within the range of $6.6 \lesssim {\rm log}~T \lesssim 6.8$.}

\section{Influence of the SED on the dynamics of the outflow}
\label{S:SEDs}

 In section 3 we presented a quasar outflow model that was fit to 
X-ray observations of \apm\ in order to constrain the column densities of the absorbers,
their ionization state and their outflow velocities.
The outflow models used to fit these X-ray data assumed that the 
incident SED is a power law with a spectral index 
of $\alpha = -1$. In this section we attempt to understand the influence of the incident SED on the dynamics of the outflow.
We vary the SED to obtain insight into the dependence of the force multiplier on the incident SED, however, we do not fit any data with variable SEDs. 
 This section extends the analysis presented in  \S3.3 from \cite{Cha09a} 
by providing an analysis of force multipliers as a function of the 
spectral changes of a \MF\ SED, while, in the  \cite{Cha09a} paper we  performed the same analysis but using a pure power-law SED.
Here we focus on the new results that are based on more realistic SEDs.

\begin{figure}
\includegraphics[width=8.6cm]{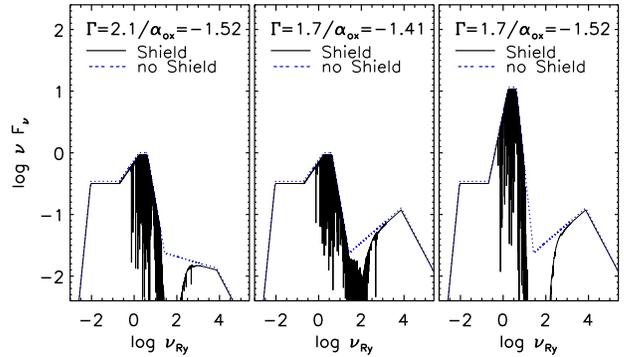}
        \centering
    \caption[Modified Mathews-Ferland SEDs used to calculate force-multipliers in  \S\ref{S:MFFM}.]{Modified Mathews-Ferland SEDs used to calculate force-multipliers in  \S\ref{S:MFFM}. The x-axis is the logarithm of frequency (in Ry units) and the y-axis is in units of the logarithm of the flux times the frequency (arbitrary units).  The left and right panels correspond to two modified versions of the Mathews-Ferland SED. In the left and right panels $\Gamma=2.1$ and $\Gamma=1.7$ with $\aox=-1.52$. The middle panel corresponds to the standard \MF\ SED, i.e.  $\Gamma=1.7$ with $\aox=-1.41$. In each panel the dotted line is the unabsorbed  SED and the solid line is the shielded SED (with $\lip _{\rm sh} \sim 2.9$).}
        \label{fig:MSED}
\end{figure}

\subsection{The dependence of the force multiplier on the photon index and \aox~ for a Mathews-Ferland SED.} \label{S:MFFM}

As suggested in Chartas et al. 2002, 2009a and Saez et al. 2009 it is plausible that the 
driving force on the high-energy absorbers is produced initially by 
X-rays. The short term variability time-scale of the X-ray BALs of \apm\ 
suggests a launching radius of $\lesssim$ $10 R_{\rm S}$
and recent studies of AGN employing the microlensing technique indicate that the X-ray emission 
region of the hot corona in AGNs is compact with a half-light radius of a few $R_{\rm S}$ and their UV regions 
are roughly a factor of ten larger \citep[e.g.,][]{Mor08,Cha09b}.
Therefore UV radiation is not expected to contribute initially at small radii to driving the X-ray absorbing outflow. 
However, as the outflowing absorber gets further away from the source the contribution of the UV photons to the driving force will increase relative to that of the X-ray photons. To investigate the driving mechanism of the wind we calculated the force multipliers for 
SEDs that extend to radio wavelengths and estimated the effect of changes in $\Gamma$.  
In Chartas et al. (2009a) we described the dependence of the force multiplier on the photon index
assuming the SED is a power-law extending from the 
UV (or 1~Ryd $\sim$13.6 eV) to hard X-rays (or $10^4$~Ryd $\sim$100 keV). 
In those calculations we assume SEDs with two different values of the power-law photon index. 
A ``soft'' ($\Gamma=2.1$) and a ``hard'' ($\Gamma=1.7$) SED.
In this work we extend our simulations to include a standard and a slightly modified version of the Mathews-Ferland SED \citep{Mat87}.
For each SED we calculated the continuum ($M_{\rm C}$) and the line ($M_{\rm L}$) components of 
the force multiplier. $M_L$ depends on an additional parameter, $t$, 
\footnote{The dimensionless optical depth is $t = n_e \sigma_T v_{\rm th}/(dv/dr)$, 
where, $n_{\rm e}$ is the electron number density,  $\sigma_{\rm T}$ is the Thomson 
cross section and  $v_{\rm th}$ is the thermal velocity of the gas. The line force 
multiplier increases with decreasing $t$.} which is commonly referred to as the 
``effective electron optical depth'' and encodes the dynamical information of the 
wind in the radiative acceleration calculation (see  Appendix~\ref{ap:FMAr}). 
For our calculations we have assumed ${\rm log}~t=-7$. 

\begin{figure}
 \includegraphics[width=8.4cm]{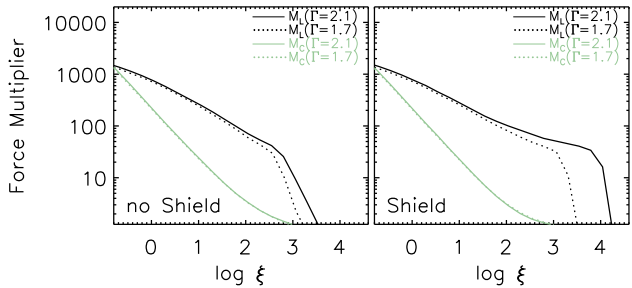}
        \centering
    \caption[The bound-free, $M_{\rm C}$, and bound-bound, $M_{\rm L}$, components of the force multiplier are shown as a function of the ionization parameter.]{The bound-free, $M_{\rm C}$, and bound-bound, $M_{\rm L}$, components of the force multiplier are shown as a function of the ionization parameter. Force multipliers are calculated for Mathews-Ferland SEDs with photon indices of $\Gamma$ = 2.1 (solid lines) and $\Gamma$ = 1.7 (dotted lines). In the left panel we have assumed no absorbing shield, whereas, in the right panel  the soft and hard SEDs have been attenuated by a warm-absorber shield with $\lnhcm =23$ and $\lip_{\rm sh}=2.9$ on its illuminated face. The SEDs used are described by the middle and right panels of Figure~\ref{fig:MSED}.} 
      \label{fig:MFFM}
\end{figure}

\begin{figure}
\includegraphics[width=8.4cm]{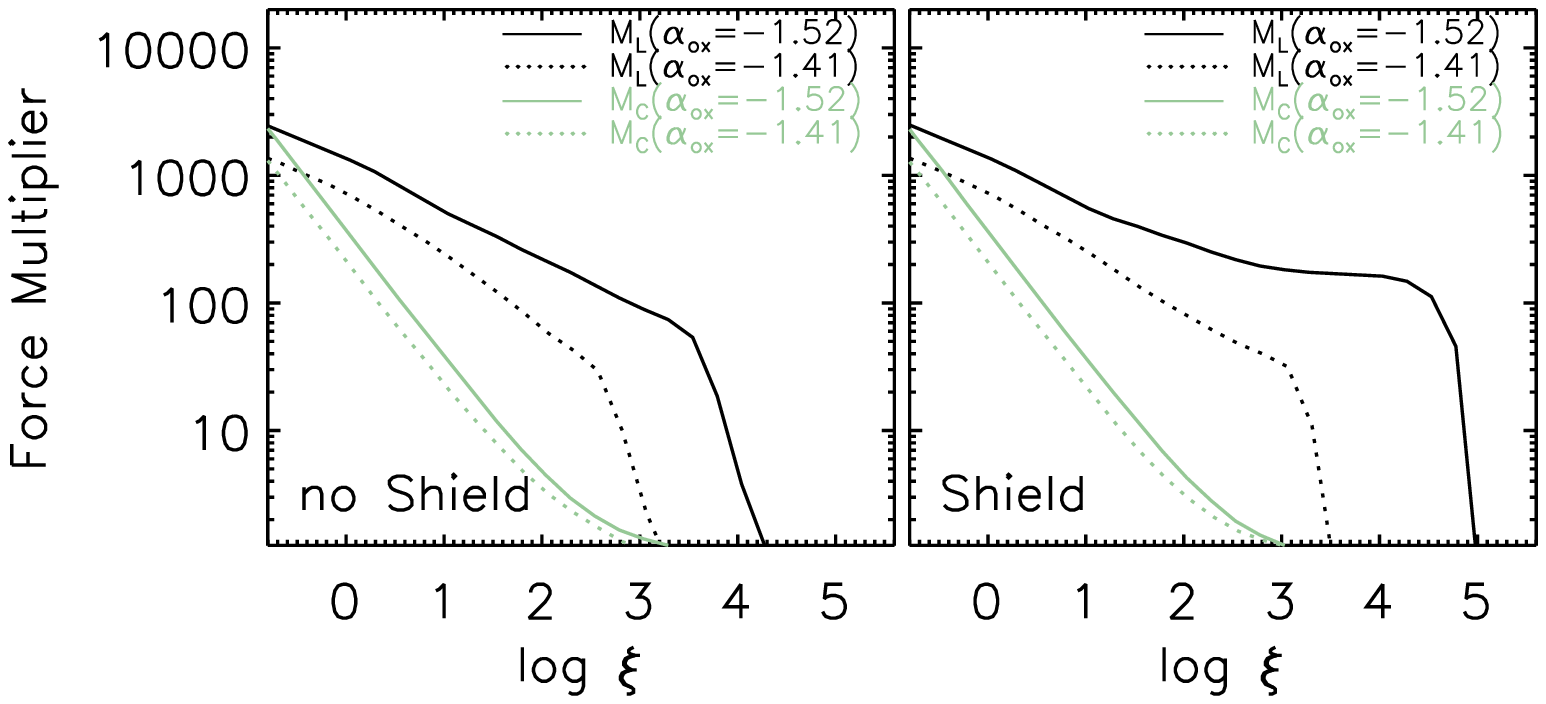}
        \centering
    \caption[The bound-free, $M_{\rm C}$, and bound-bound, $M_{\rm L}$, components of the force multiplier are shown as a function of the ionization parameter.]{The bound-free, $M_{\rm C}$, and bound-bound, $M_{\rm L}$, components of the force multiplier are shown as a function of the ionization parameter. Force multipliers are calculated for a Mathews-Ferland SED with  $\aox = -1.52$ (solid lines) and $\aox = -1.41$  (dotted lines); in both SEDs $\Gamma=1.7$. In the left panel we have assumed no absorbing shield, whereas, in the right panel  the soft and hard SEDs have been attenuated by a warm-absorber shield with $\lnhcm =23$ and $\lip_{\rm sh}=2.9$ on its illuminated face. The SEDs used are described by the middle and left panel of Figure~\ref{fig:MSED}.}
      \label{fig:MMFM}
\end{figure}

The standard Mathews-Ferland SED is characterized by $\aox=-1.41$ and by a 
\PL\ spectral index of $\alpha_X=-0.7$ ($\Gamma=1.7$)  which extends from 27 Ry ($\sim$0.36~keV) 
to 7.4$\times 10^3$ Ry ($\sim$100~keV). 
The modified Mathews-Ferland SED is similar to the standard one but with $\alpha_X=-1.1$ 
and $\aox=-1.52$ (see Figure~\ref{fig:MSED} to see the different SEDs used).
In the left panel of Figure~\ref{fig:MFFM} we show that for low ionization parameters ($\lip \lesssim 2$) the UV dominates the driving of the wind, 
therefore in this regime there is no major change in the force multipliers when we compare the soft and hard Mathews-Ferland SEDs. 
However at  $\lip \gtrsim 2$, the line force multipliers for the soft Mathews-Ferland SED case are larger than those for the hard SED case. 
This effect is produced because in the case of a Mathews-Ferland SED with a soft X-ray spectrum ($\Gamma = 2.1$)
the outflowing gas does not become very ionized at high levels of the ionization parameter
compared to the case of a \MF\ SED with a hard X-ray spectrum ($\Gamma = 1.7 $). 
We conclude that an ionizing source that is relatively soft in the X-ray band
will result in a force multiplier that remains relatively large 
even at ionization levels of the outflowing absorber of $\lip \gtrsim 3$. 
This prediction is consistent with the observed correlation of maximum outflow velocity with 
the X-ray photon index.
We modified the SEDs by including a warm-absorber shield with $\lnhcm = 23$ and  $\lip_{\rm sh} = 3.1$ (Figure~\ref{fig:MFFM}  right 
panel; see also Figure~\ref{fig:MSED} to see the shielded SED). 
The motivation of the inclusion of this absorbing shield is based on the fact that both observations
and theoretical models of radiatively driven winds suggest its existence  \citep[e.g.,][]{Mur95, Pro04}. 
We note that the observed absorption at \RF\ energies $\lesssim 2$~keV  appears to be produced by gas of low ionization parameter and could therefore be associated with this proposed shield.
The main effect of including a shield is a larger increase of the line force multipliers for the soft SED case ($\Gamma = 2.1$) compared to the increase produced by the 
hard SED case ($\Gamma =1.7$) for $\lip \gtrsim 2$. 
We also find that by including a shield we obtain significant line force multipliers at ionization levels that are larger than the ones possible with no shield present. This effect is especially important in the soft SED case.

In the process of changing the X-ray photon index from the default value of $\Gamma=1.7$ (hard-spectrum) to $\Gamma=2.1$ (soft spectrum) in the Mathews-Ferland SED, 
we also changed \aox\ from $-1.41$ to $-1.52$.
To test the sensitivity of the force multiplier to changes of the  \aox\ of the SED, we estimated the force multiplier for the case where 
\aox\ changed while the X-ray spectral slope remained constant. In order to produce this effect we increased the blue bump of  the Mathews-Ferland SED by changing the default value 
of $\aox$ from $-1.41$ to $-1.52$ (to see the SEDs used see Figure~\ref{fig:MSED}).
As shown in the left panel of Figure~\ref{fig:MMFM} the force multiplier is more sensitive to $\aox$ than $\Gamma$. This difference in sensitivity is associated with the fact that the UV covers the range where most of the absorption lines lie.  We also modified the SEDs by including an absorbing shield with $\lnhcm = 23$ and\ $\lip_{\rm sh} = 3.1$ (see Figure~\ref{fig:MSED}).  
In the right panel of Figure~\ref{fig:MMFM} we show that the effect of adding this warm-absorber shield is an increase of 
the force multiplier for soft SEDs even at high ionization levels.

We note that we used a moderate amount of shielding for the estimates shown in the left panels of 
Figures~\ref{fig:MFFM} and \ref{fig:MMFM}. 
The combination of moderate attenuation and high force multipliers in this type of shield is expected to produce high velocity outflows \citep{Che03}.  If we use an absorbing shield with a relatively low ionization level
we find that  the large absorption from the shield will attenuate a large fraction of the UV and X-ray photons, and therefore prevent the outflow from reaching high terminal velocities. 

In general, the presence of an absorbing shield will increase the differences between the line force multipliers for the soft  and hard SED cases.
These differences will be important for shields with ionization parameters in the range of  $1.0 \lesssim \lip_{\rm sh} \lesssim 1.8$ for the \PL\ SED case and 
will be important for shields with ionization parameters in the range of $1.2 \lesssim \lip_{\rm sh} \lesssim 3.1$ 
for the Mathews-Ferland SED case (see Appendix~\ref{ap:warm}). 
For  values of the ionization parameter of the shield over these boundaries, the shield will be transparent to incident radiation and the force multipliers
will be similar to ones with no shield present 
(left panels of Figures~\ref{fig:MFFM} and \ref{fig:MMFM}). 
For values of $\lip_{\rm sh} \lesssim 1.0$ the large 
absorption produced by the shield will result in a weak outflow \citep[see,][]{Che03}.

 If the $\Gamma-v_{\rm max}$ trend is confirmed with additional observations we plan to 
produce a more sophisticated model that will include more realistic kinematic, ionization and absorption properties of the outflow.  If the $\Gamma-v_{\rm max}$ relation is real we would also like to know if the changes in $\Gamma$ are accompanied by changes of  $\alpha_{ox}$. Studies of correlations between outflow properties and the SED of the ionizing source will
provide crucial information regarding the driving mechanism of the outflow. Simultaneous multiwavelength observations of sources such as \apm\ that show fast outflows will therefore be essential for the study of quasar outflows. 
We note that for the SEDs analyzed in this section the contribution to the force multiplier from highly ionized iron lines such as \FeXXV\  K$\alpha$ ($1s^{2} -1s2p$;  6.70~keV) and/or \FeXXVI\  ($1s-2p$; 6.97~keV) is relatively small. However, we observe significant blueshifts of the X-ray BALs of \apm\ indicating highly ionized material moving at near-relativistic velocities.  
One possible interpretation is that we are observing absorption from outflowing highly ionized material that 
was accelerated at a previous time when it was at a lower ionization level.  We note that a better understanding of quasar outflows will have to await for the development of more sophisticated numerical codes that combine MHD simulations of the outflowing material and the photoionization properties of the wind. Such an analysis is beyond the scope of this paper.

\section{Conclusions}

We have re-analyzed 8 long exposure X-ray observations of \apm\ (2 \chandra, 3 \xmm, and 3 \suzaku) using a new 
quasar-outflow model. 
This model is based on \cloudy\ simulations of a near-relativistic quasar outflow that were used to generate 
\xspec\ table models.  

The main conclusions from this re-analysis are: 1) In each observation we have found X-ray BALs
that have been detected at a high level of significance. Specifically, we find strong and broad absorption at rest-frame energies of \hbox{$1-4$~keV} (low-energy) and \hbox{$7-18$~keV} (high-energy).
2) We confirm that the medium producing the low-energy absorption is a nearly neutral absorber with a column density of $\lnhcm \sim 23$. The medium producing the high-energy absorption appears to be outflowing from the central source at relativistic velocities (between $0-0.7 c$) and with a range of ionization parameters ($3 \lesssim \lip  \lesssim 4$). The maximum detected projected outflow velocity of $\sim 0.7c$ constrains the angle between our line of sight and the wind direction to be $\lesssim 22^\circ$. The short time-scale variability ($\lesssim$~week in the rest-frame) of the high-energy absorption implies that X-ray absorbers are launched from distances of $\sim 10R_S$ from the central  source (where $R_S$ is the Schwarzschild radius). 3) We also confirm  trends between the  maximum projected outflow velocity ($v_{\rm max}$) with the photon index ($\Gamma$) and the intrinsic (unabsorbed) X-ray luminosity at the 2$-$10 \RF\ band; i.e. softer and more luminous X-ray spectra generate larger outflow terminal velocities.  We also find that the total column density ($\nh$) of the outflow increases with $\Gamma$. The trends found suggest that the wind becomes more powerful as the incident spectrum becomes softer. This tendency is confirmed through estimates of the efficiency of the wind. Additionally, we estimate that a significant fraction ($>$10\%) of the total bolometric energy over  the quasar's lifetime is injected into the intergalactic medium of   APM 08279+5255 in the form of kinetic  energy. 

In this work we modeled the spectrum of the central ionizing source with a power-law SED and with Mathews-Ferland SEDs 
and found that variations of the X-ray and UV parts of the SEDs
will produce important changes in the strength of the radiative driving force. 
These results support the observed trend found between the outflow velocity and X-ray photon index in \apm.
In general we find as expected that the presence of a moderate absorbing shield results in more powerful outflows  \citep{Che03}. 
Specifically, we find that shields with column densities  of $\lnhcm \sim 23$, covering an optically thin outflow,  
provide a significant increase in the driving force when their ionization parameters 
are in the range of $1.0 \lesssim \lip_{\rm sh} \lesssim 1.8$ for the case of a power-law SED and in the range of 
 $1.2 \lesssim \lip_{\rm sh} \lesssim 3.1$ for the case of a \MF\ SED.
Therefore the strength of the radiative driving force depends critically on both the column density of the shield  and its ionization level.  
 A confirmation of the results found in our simulations of quasar outflow will require new deeper \XR\  multi-wavelegth observations of quasars 
that contain clear signs of fast outflow. Such observations will allow us to correlate the properties of the outflow with properties of the SED and 
thus test our predictions.

\acknowledgments

We would like to thank Chris Done and Jane Charlton for providing 
useful information in the early stages of the development of our quasar-outflow code and Mike Eracleous for his suggestions regarding the analysis of the SEDs used in \S  \ref{S:MFFM}. We would also like to thank Gordon Garmire for his financial support
via the NASA/Smithsonian Institution grant SAO SV4-74018. GC acknowledges financial support from NASA grant NNX08AB71G.
Finally, we would like to thank the anonymous referee for his/her insightful comments that helped to improve considerably the quality of this work.

\appendix

\renewcommand\thefigure{A.\arabic{figure}}
\setcounter{figure}{0}


\section{Ionization parameter.} \label{ap:ionp}

There are two definitions of the ionization parameter that are broadly used in the scientific literature.
The first definition, which will be mostly adopted in this work,  is the ionization parameter of \cite{Tar69} given by 

\begin{equation}
\xi=\frac{L_{\rm I}}{n_H r^2}=\frac{4 \pi}{n_H}
\int_{1 \rm Ry}^{1000 \rm Ry}F_{\nu}d\nu=\frac{4 \pi F_I}{n_H},
\end{equation}

where $L_I$ is the ionizing luminosity, $F_I$ is the ionizing flux, $F_\nu$ is the incident flux,  $n_H$ is the hydrogen density, and $r$ is the source-cloud
separation. A  second definition of the ionization \citep{Dav77}
parameter is given by the ratio of photons that can ionize hydrogen to the number of hydrogen atoms in a spherical layer at a distance $r$ from an illuminating point source, i.e.,

\begin{equation}
U=\frac{Q}{4 \pi r^2 c n_H} \hspace{10pt},{\rm where}
\hspace{10pt} Q=\int_{1 \rm Ry}^{\infty}\frac{L_{\nu}}{h \nu}d\nu.
\end{equation}

We note that for a pure power law (i.e., $F_{\nu} \propto \nu^{\alpha}$), the analytic expressions to convert between 
$\xi$ and $U$ (valid for $\alpha < 0$) are,

\begin{equation}
\lip = \left\{
\begin{array}{ll}
\lU + 1.754  \hspace{20 pt} & \alpha = -1 \\
\lU + 0.914 + {\rm log} \left(-(1000^{\alpha +1}-1) \frac{\alpha}{\alpha+1} \right) \hspace{20 pt} & \alpha \neq -1
\end{array}
\right.
\end{equation}

\section{When can an absorption slab be considered thin?} \label{ap:nhma}

The fraction of flux absorbed in the rest-frame of a layer of gas depends on the ionization
state of the absorber, its column density, and the SED of the illuminating source. 
Our quasar outflow code calculates the absorption profiles 
resulting from absorption of the central source spectrum by ionized gas in an accelerated outflow. 
One of the assumptions of our quasar outflow model is that the 
absorption layer producing the observed X-ray BALs 
has a plane-parallel geometry with a thickness that is much smaller than the distance from
the central source\footnote{Let's assume a relatively faint AGN with $L_
I \sim 10^{42} \lumin$ with an absorbing layer of $\lnhcm=22$,
$n_H=10^{8} \cc$, and an ionization parameter of $\lip=3$ on the
illuminated side of the layer. Since $\xi=L_{\rm I}/(nR^2)$, $R \sim
3\times10^{15}n_{8}^{-1/2}$~cm; therefore $\Delta R/R \sim
0.3~n_{8}^{-3/2}$. Since this is an extreme case (faint AGN and
highly ionized layer), in general we expect $\Delta R/R \ll
0.1$.}.  We therefore expect any decrease in the ionization parameter 
across an absorbing layer to be predominantly due to the absorption of the flux in the layer rather than to the 1/R$^{2}$ decrease of the ionization parameter.

We will refer to a layer as optically thin if the change between the ionization
parameter on the illuminated side and the dark side of the layer
is less than 0.05~dex.\footnote{If we define $\xi_i$ and $\xi_f$ as the ionization parameters at the illuminated front and back sides of the layer, respectively, then our definition of a thin layer can be written as ${\rm log}~(\xi_f/ \xi_i) < -0.05$. In addition if we assume  that the incident flux on layer is $F_\nu$, we have that ${\rm log}~(\xi_f / \xi_i)=  \log ( \int_{\nu_0}^{\nu_1} F_\nu e^{-\tau_\nu} d\nu  / \int_{\nu_0}^{\nu_1} F_\nu d\nu) \sim {\rm log} (1-\left<\tau \right>) \sim -\left<\tau \right> / {\rm ln} 10$, where $\nu_0=1$~Ry, $\nu_1=1000$~Ry and $\left<\tau \right>$ is the average opacity through the layer between $\nu_0$ and $\nu_1$. Therefore, our definition of a thin layer is equivalent to a layer with an average opacity of   $\left<\tau \right> \lesssim 0.1$.}

     \begin{figure}
   \includegraphics[width=8cm]{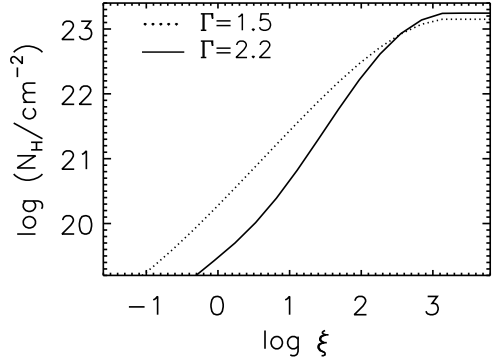}
        \centering
       \caption[Limiting column density at which a layer will be considered thin.]{Limiting column density at which a layer will be considered thin. 
       In this case we arbitrarily define a layer as thin when the ionization parameter decreases after crossing the layer less than 0.05 dex due to absorption.     
       The solid and dotted lines represent a soft spectrum power-law ($\Gamma=2.2$) and hard power-law spectrum ($\Gamma=1.5$), respectively. Both spectra are defined between 1~Ry and $10^4$~Ry}
     \label{fig:NHma}
     \end{figure}

In Figure \ref{fig:NHma} we show the column density 
of a thin layer that will produce a decrease of 0.05~dex of the ionization parameter across this
layer as a function of the ionization parameter at the front surface of the layer.
We plot this relation for power-law SEDs with a photon index of 
$\Gamma=1.5$ (dotted line) and with $\Gamma=2.2$ (solid line). 
The results show that the upper limit of the column density of an absorbing layer required for this 
layer to be considered thin, increases with ionization parameter. Soft spectra have lower
thresholds than hard spectra, especially for low values of the ionization
parameter ($\lip \lesssim 2$). However, for highly ionized media the
threshold of the soft and hard spectra are similar, in particular
for ($\lip \gtrsim 3$) the limiting column density is $\lnhcm \sim
23$.

\section{Near-relativistic quasar outflow code:  assuming an outflow with a constant $\Delta v'$ between layers} \label{ap:Dvco}
Since our library of \cloudy\ simulations was generated for a specific turbulent velocity we adjusted
the relative speed between layers $i$ and  \hbox{$i+1$} to be constant (i.e.  $\Delta v_i'=(v_{i+1}-v_i)/(1-v_i v_{i+1}/c^2)$ constant). 
Assuming $N_l$ layers, we initialize the parameters defining our profile by: 1) Deriving the velocity profile, which satisfies $v_0=v_{\rm min}$; $v_{N_l}=v_{\rm max}$ ($\Delta v_i'=\Delta v'$). 2)Deriving $\nh_i $ (i=0,..,$N_l$) through the use of $\nh_{i+1}^p-\nh_i^p=\Delta v_i/k$; this result comes from $v(\nh)=v_{\rm min}+k\nh^p$  ($k=(v_{\rm max}-v_{\rm min})/\nh_T^p$  and $\nh_T$ is the total column density of the outflow).   Using the last expression we calculate $\nh_i$ from $\nh_{i-1}$ starting from $\nh_0 = 0$.  3) Calculating the average velocity $v_{mi}=(v(r_{i+1})+v(r_i))/2$,  the column density $\Delta \nh_i=\nh_{i+1}-\nh_i$, the density $n_i=n(v_{mi})$, (i=0,..,$N_{l-1}$) and  the radial steps $\Delta r_i=\Delta \nh/n_i$ of each layer.   

We note that our assumptions are valid as long as $ \Delta v < v_{\rm turb} \sqrt{12}$ (see \S \ref{S:lays}). In our simulations we chose $\Delta v_i \sim v_{\rm turb}$ and $v_{\rm turb}=1500~\kms \sim 0.005c$ . In the case of $v_{\rm min}=0.1c$ and $v_{\rm max}=0.7c$ we would have  $N_l \sim  100$ layers.

\section{Near-relativistic quasar outflow code:  comparison with Schurch \& Done (2007)} \label{ap:Schu}

In this appendix we compare our quasar outflow code to a similar code presented in \cite{Sch07}.
The goal is to estimate the level of agreement between two independent quasar outflow codes 
and at the same time evaluate the limitations of each approach. 
We first  generated absorption profiles for a linearly accelerated outflow ($p=1$) with $v_{\rm min}=0$, $v_{\rm max}=0.3c$, 
$\nh=3 \times 10^{23}$~\cmsq , $n_0=10^{12}$~\cc, and  values of $\lip=$ 2.75, 3, 3.25, 3.5, 3.75 and 4.  In order to make a fair comparison with the results of \cite{Sch07} we 
ran a large number of simulations using the photoionization code \xstar\ 
and assumed a power-law model with $\alpha=-1.4$. 

     \begin{figure}
   \includegraphics[width=12cm]{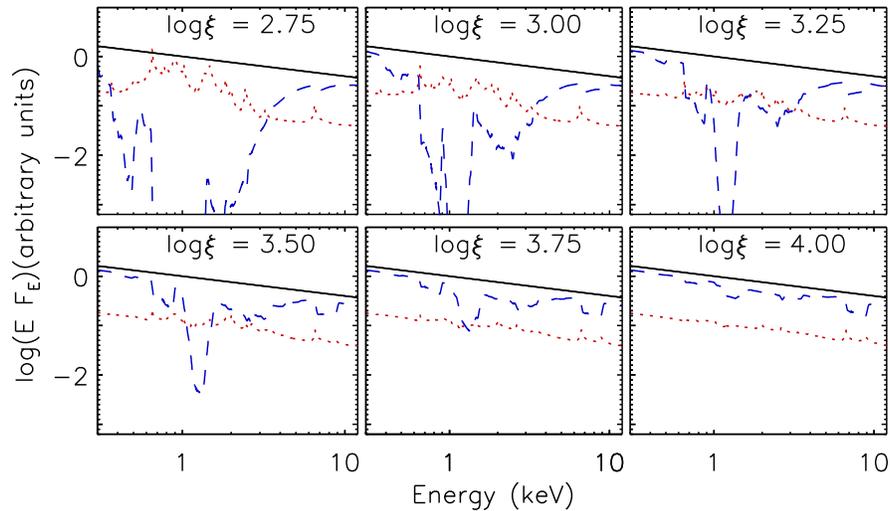}
        \centering
       \caption[0.3-13 keV spectra of the incident (full-line), transmitted (dashed-line) and emitted (dotted-line) continua for different ionization levels of the outflowing material
       simulated with our quasar outflow code.]{0.3-13 keV spectra of the incident (full-line), transmitted (dashed-line) and emitted (dotted-line) continua for different ionization levels of the outflowing material
       simulated with our quasar outflow code. In Panels 1-3 (Top, left to right) the ionization levels of the 
       outflowing material are  \lip = 2.75, 3.0 \& 3.25, respectively. In Panels 4-6 (Bottom, left to right) the ionization levels of the 
       outflowing material are  \lip = 2.75, 3.0 \& 3.25, respectively. For all these simulations the total number of layers is $N_l = 100$,
       the total column density of the outflowing material is $N_{\rm H} =3 \times 10^{23}$ \cmsq, the initial gas density is 10$^{12}$ cm$^{-3}$ 
       and the velocity of the outflow is assumed to vary linearly between 0 and 0.3 $c$. These results are in good agreement with 
       those of \cite{Sch07} as evident from a comparison between this figure and their Figure~7.}
     \label{fig:Sch1}
     \end{figure}

In each layer we subtract the absorption  an add the emission. Since \xstar\ does not include  a ``Compton scattered"  component for the emission, we add this component in a similar way as the one described in Schurch \& Done 2007. 
The Compton scattered emission component is obtained by calculating the Thomson scattered flux of each layer\footnote{The Thomson scattered flux is the incident flux multiplied by $1-e^{- \Delta N_{\rm H} \sigma_T}$ where $\Delta \nh$ is the column density of the layer $\sigma_T$ is the Thomson cross section.}, where the incident spectrum is the output from the previous layer. 
Since in the case of \cite{Sch07} the calculations were made assuming spherical symmetry of the multilayer absorber, with global covering factors equal to one, we use similar assumptions 
to obtain the emission of each layer. We calculated the emission from a spherical shell by adding the spectral contribution from each solid angle of the layer. The emission from 
each solid angle of the shell is corrected for Doppler beaming of material outflowing at a projected velocity with respect to a fixed line of sight.

         \begin{figure}
   \includegraphics[width=8cm]{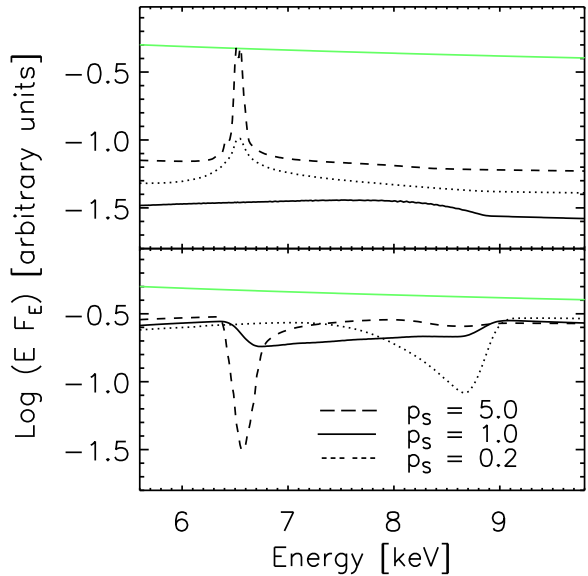}
        \centering
       \caption[Emitted (upper panel) and absorbed (lower panel) spectra in the energy range where we detect iron 
absorption ($\sim$7--9~keV) for three different values of the exponent of the $p$-type velocity profile, 
(see equation~\ref{eq:vNH}, \S\ref{S:nro})  of $p$ = 5.0, 1.0  and 0.2. ]{Emitted (upper panel) and absorbed (lower panel) spectra in the energy range where we detect iron 
absorption ($\sim$7--9~keV) for three different values of the exponent of the $p$-type velocity profile, 
(see equation~\ref{eq:vNH}, \S\ref{S:nro})  of $p$ = 5.0, 1.0  and 0.2. 
For these comparison tests we set $v_{\rm min}=0$, $v_{\rm max}=0.3c$, $\nh=3 \times 10^{23}$~\cmsq , $n_0=10^{12}$~\cc, and  $\lip= 3$. 
We find agreement between our code and that of Schurch \& Done 2007 as shown in the comparison between 
the absorption and emission profiles presented in Figure~\ref{fig:Sch2} and Figure~4 of \cite{Sch07}.}
     \label{fig:Sch2}
     \end{figure}

The spectra of the incident, absorbed and emitted continua
for each of the initial values of the ionization parameters used in these calculations of absorption line profiles
are presented in the six panels of Figure~\ref{fig:Sch1}. 
We find a remarkable similarity of Figure~\ref{fig:Sch1} with Figure~7 of Schurch \& Done 2007.
The small differences found in our results when compared to those of \cite{Sch07}  
are mainly associated with differences in the details of the calculations. 
Specifically, Schurch \& Done 2007 pass the output flux from one layer to the other
internally\footnote{In \citep{Sch07} the authors use the same photon field \xstar\ propagates internally, therefore their approach is
consistent with a single, constant density, \xstar\ run, in the absence
of an external velocity field and constant gas density throughout the
chain.}, while in our approach we perform separate photoionization runs from one layer to the next.
Possibly an important factor that contributes to the discrepancies between the two codes is that in our 
calculations we approximate the incident flux on each layer as being the unabsorbed source SED.
On the other hand, \cite{Sch07} use the absorbed SED to perform the photoionization calculations
in each layer.  
We expect our approximation of using the unabsorbed source SED as incident on all layers 
to break down at low ionization levels where absorption through a layer may
significantly change the SED. This explains why our results begin to noticeably differ from those of \cite{Sch07} at  $\lip \lesssim 3$.

In Figure~\ref{fig:Sch2} we plot the absorbed and emitted spectra in the energy range where we detect iron 
absorption ($\sim$7--9~keV) for three different values of the exponent of the $p$-type velocity profile, 
(see equation~\ref{eq:vNH}, \S\ref{S:nro})  of $p$ = 5.0, 1.0  and 0.2. 
For these comparison tests we set $v_{\rm min}=0$, $v_{\rm max}=0.3c$, $\nh=3 \times 10^{23}$~\cmsq , $n_0=10^{12}$~\cc, and  $\lip= 3$. 
We find agreement between our code and that of Schurch \& Done 2007 as shown in the comparison between 
the absorption and emission profiles presented in Figure~\ref{fig:Sch2} and Figure~4 of \cite{Sch07}. 
Our emission spectra for the case of a velocity profile with an exponent of $p=5$ 
differs from those derived by Schurch \& Done 2007 (see Figure~4 of \cite{Sch07})
mainly because in their simulations the turbulent velocity is set
to be variable in order to model fine structures in the spectra. 
In our simulations we fix the turbulent velocity in the outflow, and therefore we are not able to reproduce fine details ($\lesssim 1500$~\kms) in the spectra. 
This is not a significant problem since we are mainly interested in simulating 
the broad absorption lines detected in the X-ray spectra of BAL quasars.\footnote{A typical energy resolution of a \XR\ telescope (at energies between 1$-$10~keV) is $\sim 100$~keV, therefore for a \FeXXV\  K$\alpha$ ($1s^{2} -1s2p$;  6.70~keV) line we will not be able to resolve features with Doppler broadenings $\lesssim 4000~\kms$.}

We conclude that our quasar outflow code provides results that are very similar to
those found in the independent code of Schurch \& Done 2007 as long as the 
SED does not change significantly across the multilayer.
For our study we have mostly selected quasar outflows of highly ionized material
(i.e., $\lip$ $\gtrsim 3.0$ ) and therefore expect the SED of the central source
not to be significantly attenuated across the outflowing multilayer.  As a final remark, we note that in  our quasar outflow code the photoionization calculations at each layer use a library of preexisting runs that can be performed either with \cloudy\,  \xstar\ or any other photoionization code. This presents an advantage over the quasar outflow code \xscort\ in the sense that we can compare photoionization calculations performed with different photoionization codes which might incorporate slightly different atomic physics. Additionally, the major advantage of using our outflow code over {\sc xscort},  is a significant shortening in the computing time required for the simulation of an absorption profile.

\section{Force multiplier calculations} \label{ap:FMAr}

In this appendix we derive a formula for the force multiplier
and compare our expression to ones reported in earlier studies \citep[see e.g.,][]{Ara94, Mih84}.

\subsubsection{Force multiplier calculation}

For a medium that absorbs radiation, the force per unit volume
on the medium is

\begin{equation}
f_v=\frac{1}{c}\int \alpha_\nu F_{\nu}d \nu,
\end{equation}

where $\alpha_\nu$(cm$^{-1}$) is the absorption coefficient and $F_{\nu}$ is the incident flux on the medium.
Assuming that the flux originates from a point source with luminosity
$L_\nu$ at $r=0$ the acceleration is given by

\begin{equation}
\frac{dv(r)}{dt}=\frac{1}{4\pi r^2\rho(r)c}\int \alpha_\nu
L_{\nu}e^{-\tau_\nu}d \nu.
\end{equation}
The factor $e^{-\tau_\nu}$ corresponds to the attenuation factor at
radius r, ($\tau_\nu(r)=\int_0^r \alpha_\nu(r')dr'$).  Assuming that the central source has a mass of $M_s$ the force equation can be re-written as:

\begin{equation} \label{wineq1}
\frac{dv(r)}{dt}=\frac{n_e(r)\sigma_{\rm T}L_{\rm bol}}{4 \pi r^2 \rho(r) c}
M(r)-\frac{GM_s}{r^2}  \hspace{10pt} {\rm with} \hspace{10pt}
M(r)=\frac {\int_0^\infty \alpha_\nu(r)L_\nu
e^{-\tau_\nu(r)}d\nu}{n_e(r)\sigma_T L_{\rm bol}},
\end{equation}

where $L_{\rm bol}$ is the bolometric luminosity, $\sigma_T$ is the Thomson cross section and $M(r)$ is the force multiplier. $M(r)$ can be interpreted as
the ratio of energy absorbed by the medium to the energy absorbed
by the electrons. 

As a simple example we assume that the medium is a spherical shell with
inner and outer radii of $R$ and $R+\Delta R$, respectively. For this case
$\left<M(r)\right>=\int M(r)dr/\Delta R$; and therefore, the wind equation (\ref{wineq1}) is
approximated as

\begin{equation} \label{wineq2}
\frac{dv(r)}{dt}=\frac{ \sigma_{\rm T}L_{\rm bol}}{1.2
m_{\rm p} 4 \pi r^2  c} \left<M(r)\right>-\frac{GM_s}{r^2}
\end{equation}

In this last stage we have assumed a high ionization state for the gas at solar metallicities and therefore $\rho_r/n_e(r) \sim 1.2 m_{\rm p}$, where $m_{\rm p}$ is the mass of the proton.
Assuming that $v(R_{\rm min})=0$ the solution of equation~(\ref{wineq2}) is

\begin{equation} \label{wineq3}
v(r)=\sqrt{2GM_s\left(\frac{L_{\rm bol}}{1.2 L_{\rm
Edd}}\left<M\right> -1\right)\left(\frac{1}{r_{\rm
min}}-\frac{1}{r}\right)}
\end{equation}

where $L_{\rm Edd}$ is the Eddington Luminosity (i.e. $L_{\rm Edd}=\frac{4\pi cG \mu m_p M_s}{\sigma_T}$). 

To calculate the force multiplier of a line we start from equation
(\ref{wineq1}).  Let's assume that the absorption line has a constant opacity, frequency $\nu_l$ and a width $\Delta \nu_{\rm th}
=\nu_l v_{\rm th}/c$ (where $v_{\rm th}$ is the thermal speed of
the absorbing atoms\footnote{Notice that we are using just thermal broadening in the force multiplier calculations. We are not using turbulent velocity in these calculations. The use of turbulent velocities in this work is to simulate the effect of the acceleration between layers in the obtention of absorption profiles using our multilayered nearly relativistic outflow model \S \ref{S:nro}. Moreover, by using only thermal broadening 
in our calculations, we can directly compare our results to those
of other authors. }). The characteristic distances of radiative
interactions are of the order of $l\sim v_{\rm th}/(\nabla v)$.
Hence for radially streaming radiation in an expanding medium
the optical thickness is,

\begin{equation}
\tau_l=\alpha_l v_{\rm th}/(dv/dr)
\end{equation}

where $\alpha_l$  is the absorption coefficient of the line. We define $\eta_l=\frac{\alpha_l}{n_e \sigma_T}$ and introduce an
equivalent electron optical depth scale

\begin{equation}
t= \tau_l/\eta_l= n_e \sigma_T  v_{\rm th}/(dv/dr).
\end{equation}

Therefore the mean force multiplier is $\left<M(r)\right>=\int M(r)dr/l$,
i.e.,

\begin{equation}
\left<M(r)\right>=\frac {\int_r^{r+l} M(r)dr}{l}=\frac{L_{\nu_l}
\Delta \nu_{th} (1-e^{-\tau_l})}{n_e \sigma_T l~L_{\rm bol}}.
\end{equation}

Summing up over all absorption lines we obtain

\begin{equation} \label{eq:Fmul}
M(t)=\frac{1}{L_{\rm bol}} \sum_l \frac{L_{\nu_l} \Delta \nu_{th,l}
(1-e^{-\eta_l t})}{t} \approx \frac{1}{L_{\rm bol}} \sum_l L_{\nu_l} \Delta
\nu_{th,l} ~{\rm min} \left(\eta_l,\frac{1}{t} \right).
\end{equation}

To obtain $\alpha_l$ we use equation 1.78 from \cite{Ryb85} i.e.,

\begin{equation}
\alpha_l=\frac{h \nu_l}{4 \pi} n_l
B_{lu}\left(1-\frac{g_ln_u}{g_un_l}\right) \phi(\nu).
\end{equation}

Assuming that the absorption line has a square-topped profile of width
$\Delta \nu_{\rm th,l}$, $\phi=\frac{1}{\Delta v_{\rm th,l}}$, and
using $B_{lu}=\frac{4 \pi^2 e^2}{h \nu_lmc}f_{l}$, where $f_l$ is
the oscillator strength we obtain

\begin{equation}  \label{eq:alpl}
\alpha_l=\frac{\pi e^2}{m_e c} g_lf_l
\frac{(n_l/g_l-n_u/g_u)}{\Delta \nu_{th,l}} .
\end{equation}

Inserting equation \ref{eq:alpl} into \ref{eq:Fmul}  we reproduce the result of \cite{Ara94}, i.e.,

\begin{equation}
M(t)=\frac{1}{L_{\rm bol}} \sum_l L_{\nu_l} \Delta \nu_{th,l} \frac{
1-e^{-\eta_l t}}{t} \hspace{20pt} {\rm with} \hspace{20pt}
\eta_l=\frac{\pi e^2}{m_e c} gf_l \frac{(n_l/g_l-n_u/g_u)}{n_e
\sigma_T \Delta \nu_{th,l}} .
\end{equation}

\subsection{Comparison with other results.}

The force multiplier mainly depends on the SED of the source, the composition of the gas, and its ionization parameter. 
In general a soft spectrum irradiating a gas that is nearly neutral (i.e. low ionization parameter) will 
strongly accelerate the gas and the force multiplier will be relatively large.

  \begin{figure}
   \includegraphics[width=8cm]{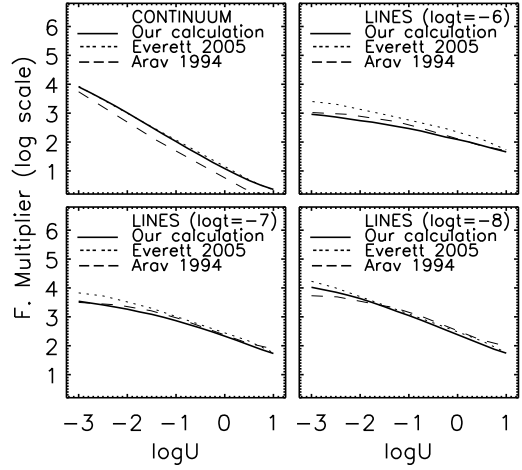}
        \centering
\caption[A comparison between our force multiplier calculations (solid
lines) and those of Everett 2005 (dotted lines) and Arav et al. 1994 (dashed lines).]{A comparison between our force multiplier calculations (solid
lines) and those of Everett 2005 (dotted lines) and Arav et al. 1994 (dashed lines). The
upper-left panel shows the continuum force multiplier as a function of the
ionization parameter $U$. The other three panels show the line force multiplier
as function of $U$ for three different values of $t$.}
\label{fig:FMAr}
\end{figure}

When the gas is highly ionized it absorbs less radiation 
and is subject to a weaker radiative driving force  (i.e., the force multiplier is relatively small).

Figure~\ref{fig:FMAr} shows the estimated force multiplier assuming a Mathews-Ferland SED. Figure~\ref{fig:FMAr} has four panels, the upper-left panel represents the continuum component of the force multiplier, mainly produced by bound-bound transitions. The other three panels in Figure~\ref{fig:FMAr} show the contribution to the force multiplier from bound-free transitions for three different values of the dimensionless factor $t$; ${\rm log}~t=$~$-6$, $-7$ and $-8$. The line force multipliers increase as $t$ decreases. This increase is noticeable when the medium is not highly ionized  ($\lU \lesssim -1$; $\lip \lesssim 0.3$).  Our  calculations are in good agreement with those performed by \cite{Ara94,Eve05}.

\section{Force multipliers calculations of material illuminated by absorbed power-laws and Mathews-Ferland SED} \label{ap:warm}

A warm absorber shielding the outflow may increase the effectiveness of radiation driving. In this appendix we use the same 
unabsorbed SEDs used in \S \ref{S:SEDs} to estimate the amount of shielding required to 
significantly increase the radiative driving force. 
Additionally, we estimate the differences between the radiative forces produced by 
a soft incident spectrum and a hard incident spectrum 
for different ionization levels of the absorbing outflow.
As in \S \ref{S:SEDs} we assume that the warm absorber shielding the outflow has $\lnhcm =23$.

 \begin{figure}
   \includegraphics[width=12.8cm]{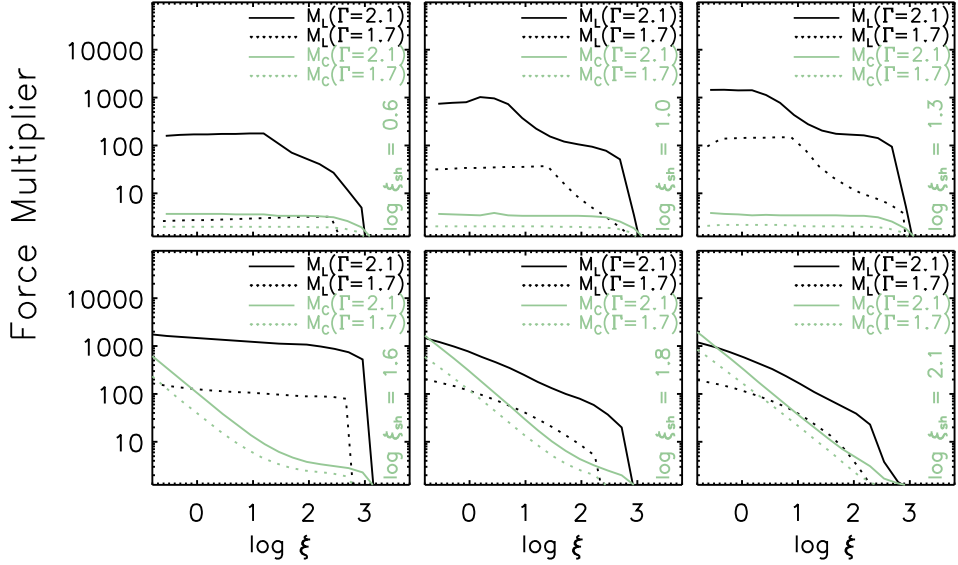}
        \centering
\caption[The bound-free, $M_{\rm C}$, and bound-bound, $M_{\rm L}$, components of the force multiplier are shown as a function of the ionization parameter (power-law SEDs).]{The bound-free, $M_{\rm C}$, and bound-bound, $M_{\rm L}$, components of the force multiplier are shown as a function of the ionization parameter. Force multipliers are calculated for a power-law SEDs with photon indices of $\Gamma$ = 2.1 (solid lines) and $\Gamma$ = 1.7 (dotted lines). In the upper-left, upper-right, lower-left and lower-right panel we have assumed absorbing shield where the soft and hard SEDs have been attenuated by a  warm-absorber shield with $\lnhcm =23$ and $\lip_{\rm sh}=$~0.6, 1.0, 1.3, 1.6, 1.8 and 2.1 respectively.}
\label{fig:P4FM}
\end{figure}

  \begin{figure}
   \includegraphics[width=12.8cm]{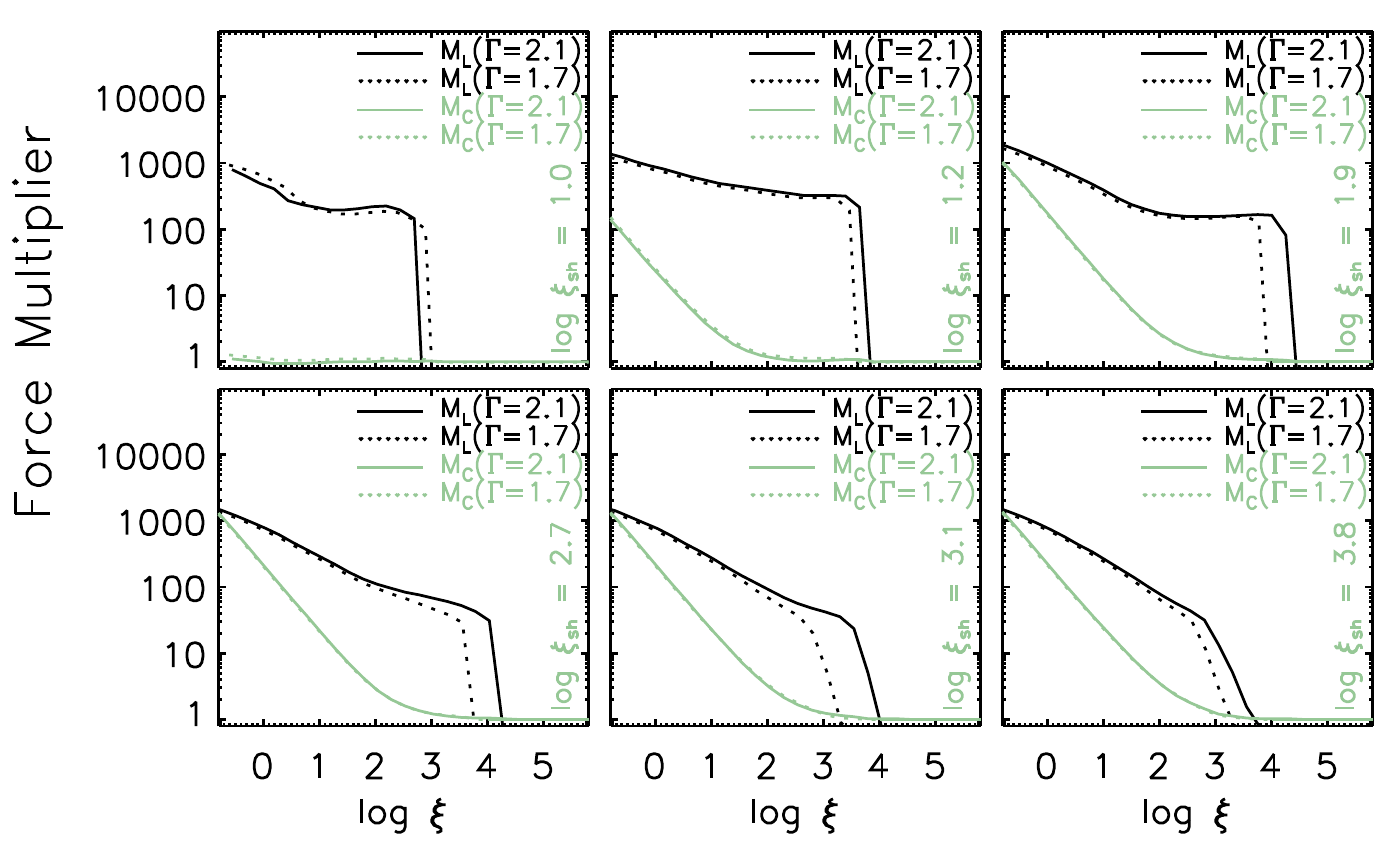}
        \centering
\caption[The bound-free, $M_{\rm C}$, and bound-bound, $M_{\rm L}$, components of the force multiplier are shown as a function of the ionization parameter (\MF\ SEDS, 1).]{The bound-free, $M_{\rm C}$, and bound-bound, $M_{\rm L}$, components of the force multiplier are shown as a function of the ionization parameter. Force multipliers are calculated for \MF\ SEDs with X-ray photon indices of $\Gamma=2.1$ ($\aox=-1.52$; solid lines) and $\Gamma$ = 1.7 ($\aox=-1.41$ dotted lines). In the upper-left, upper-right, lower-left and lower-right panels we have assumed that the SEDs have been attenuated by shields
with $\lnhcm =23$ and $\lip_{\rm sh}=$~1.0, 1.2, 1.9, 2.7, 3.1 and 3.8, respectively. 
The default parameters of a Mathews-Ferland SED are  $\Gamma=1.7$  and $\aox=-1.41$.}
\label{fig:M4FM}
\end{figure}

  \begin{figure}
   \includegraphics[width=12.8cm]{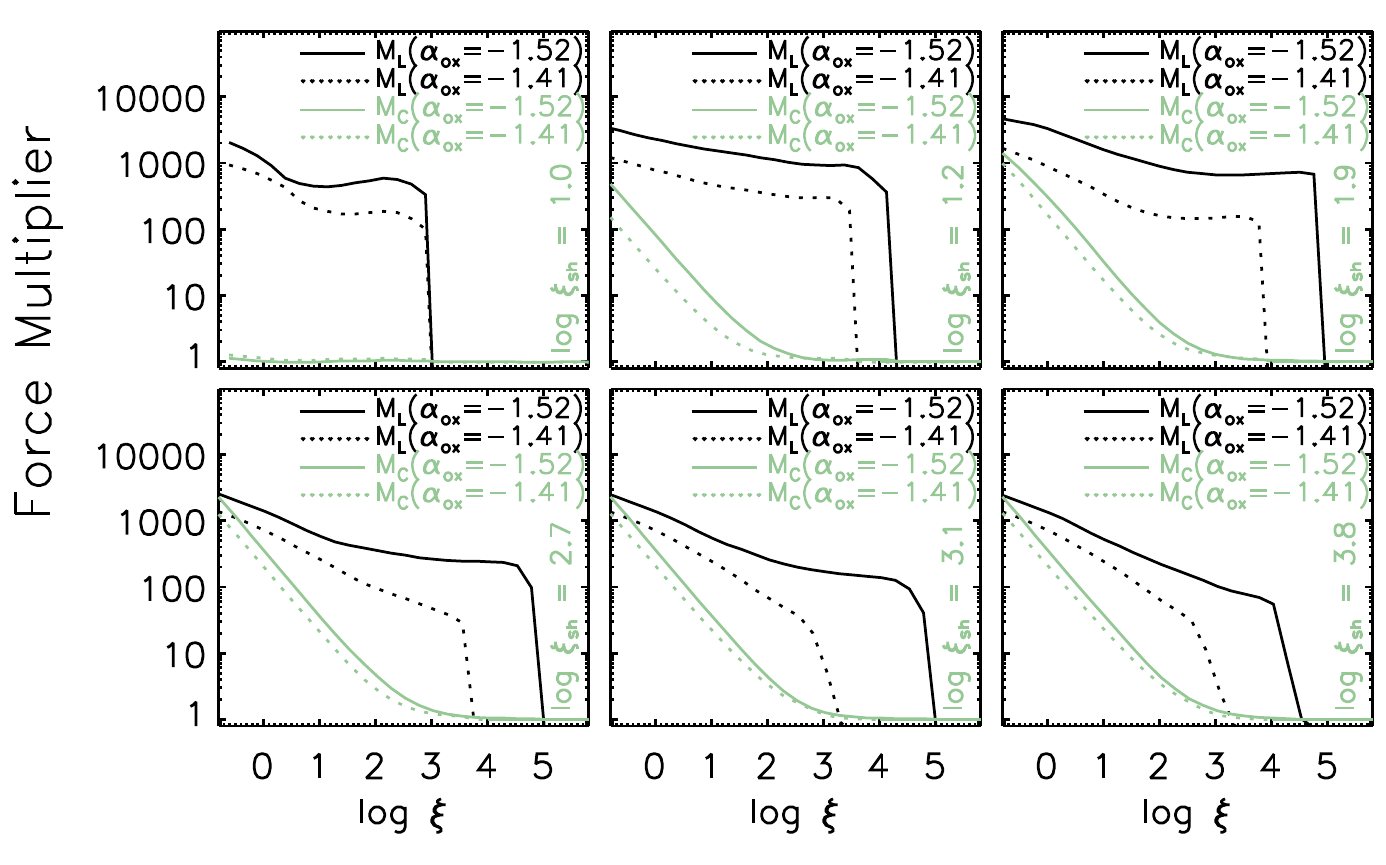}
        \centering
\caption[The bound-free, $M_{\rm C}$, and bound-bound, $M_{\rm L}$, components of the force multiplier are shown as a function of the ionization parameter (\MF\ SEDs, 2).]{The bound-free, $M_{\rm C}$, and bound-bound, $M_{\rm L}$, components of the force multiplier are shown as a function of the ionization parameter. Force multipliers are calculated for \MF\ SEDs with X-ray photon indices of $\aox=-1.52$ ($\Gamma$ = 1.7; solid lines) and $\aox=-1.41$ ($\Gamma$ = 1.7; dotted lines). In the upper-left, upper-right, lower-left and lower-right panels we have assumed that the SEDs have been attenuated by absorbing shields
with $\lnhcm =23$ and $\lip_{\rm sh}=$~1.0, 1.2, 1.9, 2.7, 3.1 and 3.8, respectively. 
The default parameters of a Mathews-Ferland SED are  $\Gamma=1.7$  and $\aox=-1.41$.}
\label{fig:MM4F}
\end{figure}

In Figure~\ref{fig:P4FM} we have plotted the bound-free, $M_C$, and the bound-bound, $M_L$, components of the force multiplier as a 
function of the ionization parameter for soft ($\Gamma=2.1$) and hard ($\Gamma=1.7$) absorbed power-law SEDs. 
In the upper-left, upper-right, lower-left and lower-right panel we have assumed absorbing shields with $\lnhcm=23$ and $\lip_{\rm sh}$~=~0.6, 1.0, 1.3, 1.6, 1.8 and 2.1, respectively.  
In Figure~\ref{fig:P4FM} we
show that $M_{\rm L}$ is significantly larger for incident SEDs with soft spectra (i.e., ${\Gamma = 2.1}$) 
than for incident SEDs with hard spectra (i.e., ${\Gamma = 1.7}$) especially at low ionization levels of the shielding gas. For values of $\lip_{\rm sh} \lesssim 0.6$  we do not expect strong radiative driving given the heavy attenuation of the continuum
 \citep[see e.g.,][]{Che03}. 
 We also note that the force multipliers $M_C$  and $M_L$ are significantly reduced for low ionization levels of the shield as shown in 
Figure~\ref{fig:P4FM}. Additionally, when  $\lip_{\rm sh} \gtrsim 2.0$ the shielding becomes transparent to the radiation and therefore the force multipliers are similar to 
the ones in the unabsorbed case. Finally from Figure~\ref{fig:P4FM} we see that when the ionization parameter of the shield lies in the range of $1.0 \lesssim \lip_{\rm sh} \lesssim 1.8$ the line force multiplier of the outflowing material is greater with the presence of an absorbing shield than without one.

In Figure~\ref{fig:M4FM} we have plotted $M_C$, and $M_L$ as a function of the ionization parameter for soft ($\Gamma=2.1$; $\aox=-1.51$) and hard ($\Gamma=1.7$; $\aox=-1.41$) absorbed Mathews-Ferland SEDs. 
In the upper-left, upper-right, lower-left and lower-right panels we assume shields that attenuate the SEDs
with $\lnhcm=23$ and $\lip_{\rm sh}$~=~1.0, 1.2, 1.9, 2.7, 3.1 and 3.8, respectively.
In the case of a \MF\ SED incident on a shield with an ionization parameter in the range of  $1.2 \lesssim \lip_{\rm sh} \lesssim 3.8$
 we find that the line force multiplier $M_L$ for soft SEDs (i.e.,  $\Gamma=2.1$) extends to
 larger values of the ionization parameter  of the outflowing material compared to hard SEDs (i.e.,  $\Gamma=1.7$).

In Figure~\ref{fig:MM4F} we have plotted  $M_C$, and $M_L$ as a function of the ionization parameter of the outflow for soft ($\Gamma=1.7$; $\aox=-1.51$) and hard ($\Gamma=1.7$; $\aox=-1.41$) absorbed modified Mathews-Ferland SEDs. In the upper-left, upper-right, lower-left and lower-right panels we 
assume shields that attenuate the SEDs with $\lnhcm=23$ and $\lip_{\rm sh}$~=~1.0, 1.2, 1.9, 2.7, 3.1 and 3.8, respectively.
In the case of a modified \MF\ SED incident on a shield with an ionization parameter in the range of 
$1.0 \lesssim \lip_{\rm sh} \lesssim 3.9$, 
we find that the line force multiplier $M_L$ is 
significantly larger for soft SEDs (i.e., $\aox = -1.52$) than for hard SEDs (i.e., $\aox = -1.41$).
We note that the line force multiplier for the case of a soft \MF\ SED incident on a shield
with an ionization parameter of $\lip_{\rm sh}$=1.9 has a value larger than $\sim$1000 
for ionization parameters of the outflowing absorber in the range of  $0 \lesssim \lip_{\rm } \lesssim 5$. For a shield with $\lip_{\rm sh} \lesssim 1$ we predict that the continuum becomes too attenuated to allow effective radiative driving. 
When  $\lip_{\rm sh} \gtrsim 3.1$ the shield becomes transparent to the incident radiation and therefore the force multipliers are similar to the ones 
estimated for the unabsorbed case (see lower right panels of Figures~\ref{fig:M4FM} and \ref{fig:MM4F}).

\begin{figure}
   \includegraphics[width=10cm]{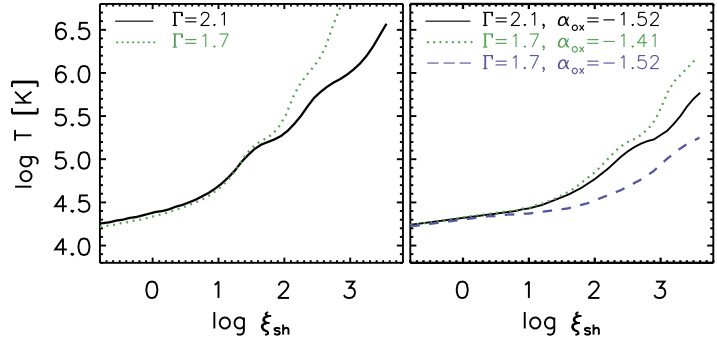}
        \centering
\caption[Temperature of a warm absorber (shield) as a function of its ionization parameter (illuminated side) for a warm absorber with $\lnhcm=23$.]{Temperature of a warm absorber (shield) as a function of its ionization parameter (illuminated side) for a warm absorber with $\lnhcm=23$. In the left panel 
the illuminating source is a power law with $\Gamma = 2.1$ (solid line) and $\Gamma = 1.7$ (dotted line). In the right panel
the illuminating source is a modified \MF\ SED with $\Gamma = 2.1$ and $\aox=-1.52$ (solid line), $\Gamma = 1.7$  and $\aox=-1.41$ (dotted line), and $\Gamma = 1.7$  and $\aox=-1.52$ (dashed line).}
\label{fig:Tshi}
\end{figure}

In Figure~\ref{fig:Tshi}  we plot the temperature of an absorbing shield as a function
of its ionization parameter (on its illuminated side) for a shield with $\lnhcm=23$.
From the analysis in the last three paragraphs and from Figure~\ref{fig:Tshi} we conclude that the most powerful winds are probably  produced at temperatures\footnote{These are temperatures at the illuminated face of the shield (output from \cloudy).} in the range $4.5 \lesssim {\rm log}~T \lesssim 5.2$.  This occurs for power-law SEDs incident on shields with $1.0 \lesssim \lip_{\rm sh}  \lesssim 1.8 $ and for Mathews-Ferland SEDs incident on shields with $1.2 \lesssim \lip_{\rm sh}  \lesssim 3.1$.

\end{document}